\documentclass{ctptex}

\usepackage[ctptex]{graphicx}
\usepackage{graphicx,subfigure,epsfig}

\usepackage{epstopdf}

\begin{document}
\begin{CJK*}{GBK}{song}

\title{The ringing of quantum corrected Schwarzschild black hole with GUP\thanks{Project supported by the National Natural Science Foundation of China (Grant Nos. 11465006, Grant Nos.11565009) and the Special Research Fund for Natural Science of Guizhou University (Grant No. X2020068). }}


\author{Yujia Xing$^{1}$,\ Yi Yang$^{1}$,\ Dong Liu$^{1}$, \ Zheng-Wen Long  $^{1}$\thanks{Corresponding author. E-mail:zwlong@gzu.edu.cn} and \ Zhaoyi Xu $^{1}$\thanks{Corresponding author. E-mail:zyxu@gzu.edu.cn}\\
$^{1}${College of Physics, Guizhou University, Guiyang, 550025, China}}  

\date{\today}
\maketitle

\begin{abstract}
Schwarzschild black holes with quantum corrections are studied under scalar field perturbations and electromagnetic field perturbations to analyze the effect of the correction term on the potential function and quasinormal mode (QNM). In classical general relativity, spacetime is continuous and there is no existence of the so-called minimal length. The introduction of the correction items of the generalized uncertainty principle (GUP), the parameter $\beta$, can change the singularity structure of the black hole gauge and may lead to discretization in time and space. We apply the sixth-order WKB method to approximate the QNM of Schwarzschild black holes with quantum corrections and perform numerical analysis to derive the results of the method. Also, we find that the effective potential and QNM in scalar fields are larger than those in electromagnetic fields.
\end{abstract}



\section{Introduction}
\label{sec:intro}
\indent Black holes are one of the most important predictions of Einstein's general theory of relativity, and in order to more fully understand the physical properties of black holes, physicists have constructed different black hole solutions \cite{Arias1,Arias2}. The shadow of the M87 \cite{M871,M872,M873} massive black hole, surrounded by a crescent-shaped aperture, was first obtained by the Event Horizon Telescope. Therefore, the black hole at the heart of the M87 elliptical galaxy has also become a key object of observation and analysis for many physicists. The LIGO-VIRGO collaboration \cite{ligo1,ligo2,ligo3,ligo4} has successfully detected gravitational waves from the combination of a black hole and a neutron star, and this discovery is extremely important for the study of black holes, in other words through the observation of gravitational waves we can fully understand the nature of the event horizon of black holes. Gravitational waves from some dense objects produce echoes \cite{Yang1,Yang2,Yang3,5,6,7,8,X1,X2,X3,X4} during propagation, which makes the echoes inseparably linked to the unique properties of dense objects, and one uses this signal to analyze the inherent properties of dense objects. It is also possible to observe gravitational wave echoes through the later stages of the QNM ringing.\\
\indent Chandrasekhar \cite{1000} made a prominent contribution to the QNM in black holes, especially the effect of the QNM on the external perturbation of the black hole, the essence of which is that there is a QNM ringing at the late stage of the initial pulse for the outside of the black hole event horizon. QNM \cite{1,2,3,4,q1,q2,q3,q4,q5,q6,q7,q8,q9} is the vibrational frequency that is available at the pure outgoing wave at infinity and the pure incident wave in the event horizon with free vibration. QNM has been widely applied in recent years, including QNM in dark matter halos \cite{Liu,l1,h3}, QNM in charged black holes with electromagnetic field perturbations \cite{17}, QNM in the multi-dimensional Einstein-power-Maxwell case with scalar field perturbations in the spacetime context of black holes \cite{18}, perturbation studies of black holes to solve spectral related problems \cite{19}, the relationship between QNM and shading with quantum correcting \cite{20}, calculation of slow revolving black holes with dynamical Chern-Simons gravity using QNM\cite{h1}, and the accurate solutions of Kerr black hole perturbations by using QNM\cite{h2,h4}. \\
\indent The solution of a black hole has a singularity, there is a singularity that would cause general relativity to fail at that point. Therefore, physicists have made a lot of research to conclude that the GUP can be used to correct the black hole degree gauge \cite{Gupta1,Gupta2,Gupta3,Gupta4,Gupta5}, and have achieved important results. In $1968$, Bardeern presented the regular black hole solutions \cite{100}. In \cite{Kazakov}, Kazakov and Solodukhin used quantum correction terms to deform the Schwarzschild solution to avoid singularities. The QNM of the corrected Schwarzschild solution has been analyzed in \cite{Konoplya}. In \cite{Hajebrahimi}, electrons can be emitted across the potential barrier into the vacuum and thus quantum corrected Schwarzschild black holes embodying Hawking radiation. In view of the importance of the QNM to reflect the relevant properties near the apparent horizon of the black hole event, we will study the QNM of the black hole with GUP correction considered in this paper.  \\
\indent This paper is structured as follows. In Sec.\ref{sec:2} \quad we use the equations of the quantum-corrected metric in the presence of scalar and electromagnetic field perturbations to derive the corresponding effective potential bases and draw an image of the effective potential. In Sec.\ref{sec:3} \quad numerical calculation by WKB method to obtain the frequency of QNM. In Sec.\ref{sec:4} \quad the conclusions of this study were drawn.\\

\section{The methods}\label{sec:2}
\subsection{Scalar and electromagnetic field perturbations in quantum-corrected Schwarzschild black hole}
\label{startsample}
\indent The uncertainty principle of black holes suggests a relationship by which the uncertainty principle at the microscopic scale is linked to black holes at the macroscopic scale. This correspondence suggests the possible existence of black holes with masses below the Planck scale, but with order radius below the Compton scale rather than the Schwarzschild scale. To eliminate the singularity of the black hole metric, we use GUP to quantum correct the Schwarzschild black hole to analyze the QNM, as the Schwarzschild radius $r_\text{h}$ is transformed into the GUP radius $r_\text{H}$ \cite{hou}, considering the GUP
\begin{eqnarray}
\Delta x> \frac{h}{\Delta p}+(\frac{\alpha l_\text{P1}^{2}}{\hbar})\Delta p,
\label{equ0}
\end{eqnarray} 
where $\alpha$ is postive dimendionless parameter, $l_\text{P1}$ is the Planck length, $l_\text{P1}=\sqrt{\frac{\hbar G}{c^{3}}}$, $M_\text{P1}$ is Planck mass, $M_\text{P1}=\sqrt{\frac{\hbar c}{G}}$, $M_\text{C}$ is Compton mass, and considering $\Delta x\rightarrow R$, $\Delta p\rightarrow cM$, then becomes
\begin{eqnarray}
R> {R}'_\text{C}\equiv \frac{\hbar}{M_\text{C}}+\frac{\alpha GM}{c^{2}}=\frac{\hbar}{M_\text{C}}\left [ 1+\alpha(\frac{M}{M_\text{P1}})^{2}\right ].
\label{equ01}
\end{eqnarray}
GUP affects the size of the black hole horizon, then Eq.(\ref{equ01}) is written as
\begin{eqnarray}
R> {R}'_\text{S}= \frac{\alpha GM}{c^{2}}\left [ 1+\frac{1}{\alpha}(\frac{M_\text{P1}}{M})^{2}\right ].
\label{equ02}
\end{eqnarray}
The free constant of Eq.(\ref{equ01}) is related to the first term, then Eq.(\ref{equ01}) and (\ref{equ02}) are written as

\begin{eqnarray}
{R}'_\text{C}=\frac{\beta  \hbar}{M_\text{C}} \left [ 1+\frac{2}{\beta}(\frac{M}{M_\text{P1}})^{2}\right ]
\label{equ03}
\end{eqnarray}
and
\begin{eqnarray}
{R}'_\text{S}= \frac{2GM}{c^{2}}\left [1+\frac{\beta}{2}(\frac{M_\text{P1}}{M})^{2}\right ]
\label{equ04}
\end{eqnarray}
where $\beta$ is postive dimendionless parameter, $h=c=G=1$.

The corrected Schwarzschild black hole:
\begin{eqnarray}
\text{d}s^{2}=f(r)\text{d}t^{2}-\frac{1}{f(r)} \text{d} r^{2}-r^{2}(\text{d}\theta ^{2}+\sin^{2}\theta \text{d}\phi ^{2}),
\label{equ1}
\end{eqnarray}
where
\begin{eqnarray}
f(r)=1-\frac{2}{M_\text{P1}^{2}}\frac{M}{r}(1+\frac{\beta }{2}\frac{M_\text{P1}^{2}}{M^{2}}),
\label{equ2}
\end{eqnarray}
and $r_{\text{H}}$ is written as
\begin{eqnarray}
r_\text{H}=\frac{2}{M_\text{P1}^{2}}(\frac{M^{2}+\frac{\beta }{2}M_\text{P1}^{2}}{M})=R_\text{S}'
,
\label{equ3}
\end{eqnarray}
with Schwarzschild black hole horizon $r_{\text{h}}$=2$M$.

\indent In this study, we focus on the analysis of spacetime in the context of Schwarzschild black holes after quantum corrections in scalar and electromagnetic fields. Then the radial equation and the effective potential are derived.\\
\indent For the massless scalar field :
\begin{eqnarray}
\frac{1}{\sqrt{-g}}{\partial \mu}(\sqrt{-g}g^{\mu\nu} {\partial \nu}\Phi) =0.
\label{equ4}
\end{eqnarray}
\indent For the electromagnetic field equation :
\begin{equation} 
\frac{1}{\sqrt{-g}}{\partial \nu }(F_{\rho\sigma}g^{\rho\mu}g^{\sigma\nu}\sqrt{-g})=0,       
\label{equ5}                
\end{equation}
where $ F_{\rho\sigma}={\partial \rho}A^\sigma-{\partial \sigma}A^\rho $, $A_\nu$ is an electromagnetic four-potential.

\indent The tortoise coordinate is expressed as:
\begin{eqnarray}
\text{d}r_{\ast}=\frac{\text{d}r}{f(r)}.
\label{equ6}
\end{eqnarray}
\indent By splitting the variances Eqs.(\ref{equ4}) and (\ref{equ5}), the equation typically takes the Schrodinger form:
\begin{equation}
-\frac{\text{d}^2\Psi}{\text{d}{r^2_*}}+V(r) \Psi=\omega ^{2}\Psi.  
\label{equ7}                                                   
\end{equation}
\indent The effective potentials under the scalar and electromagnetic fields :
\begin{equation}
V_{\text{eff}}=\frac{f(r)}{r}\frac{\text{d}f(r)}{\text{d}r}+\frac{f(r)l(l+1)}{r^{2}},  
\label{equ8}                                                   
\end{equation}
\begin{equation}
V_{\text{em}}=\frac{f(r)l(l+1)}{r^{2}}.  
\label{equ9}                                                   
\end{equation}
\begin{figure*}[t!]
\centering
\subfigure[]{ 
\label{fig:b}     
\includegraphics[width=0.3\columnwidth]{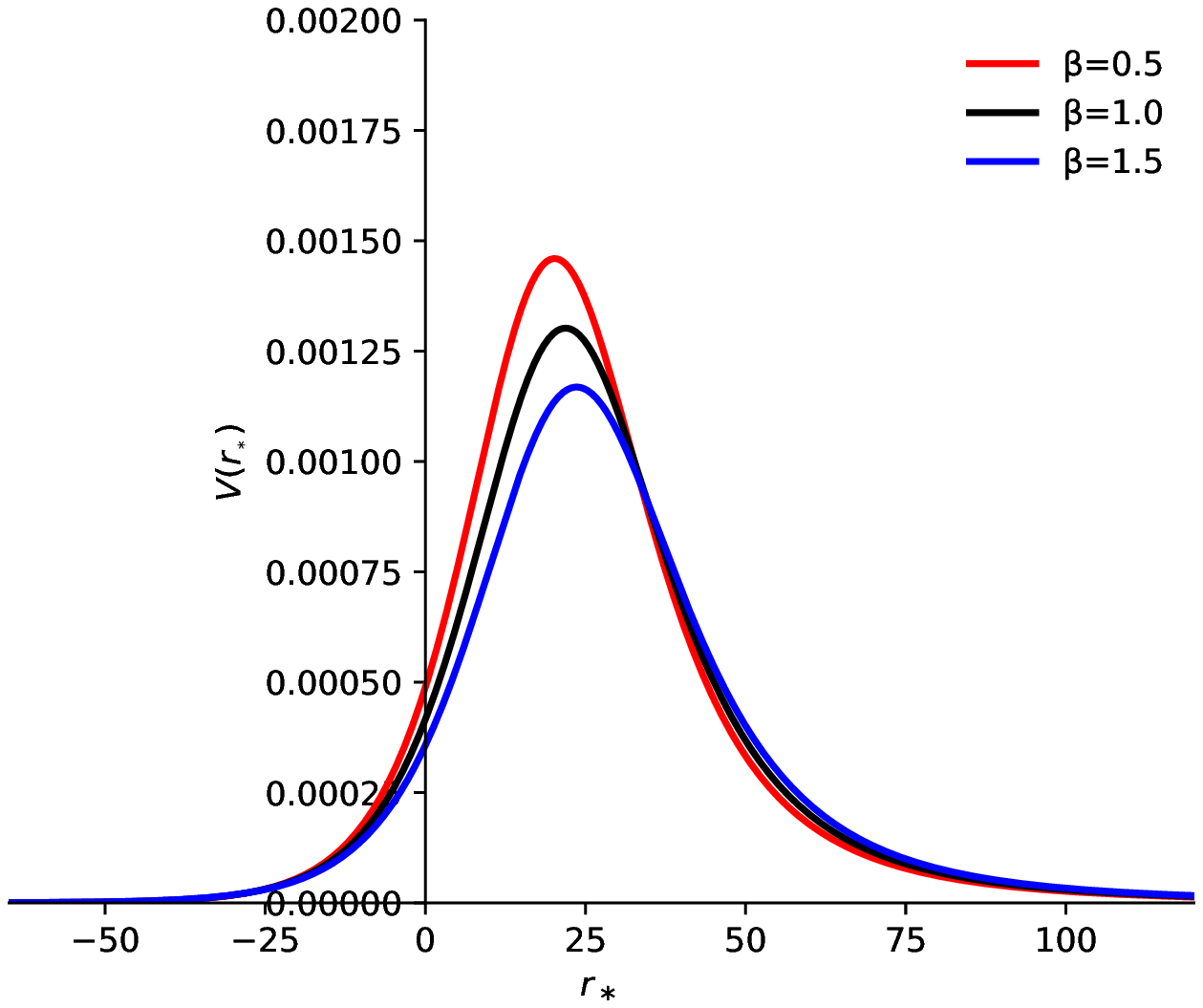}     
} 
\subfigure[]{ 
\label{fig:b}     
\includegraphics[width=0.3\columnwidth]{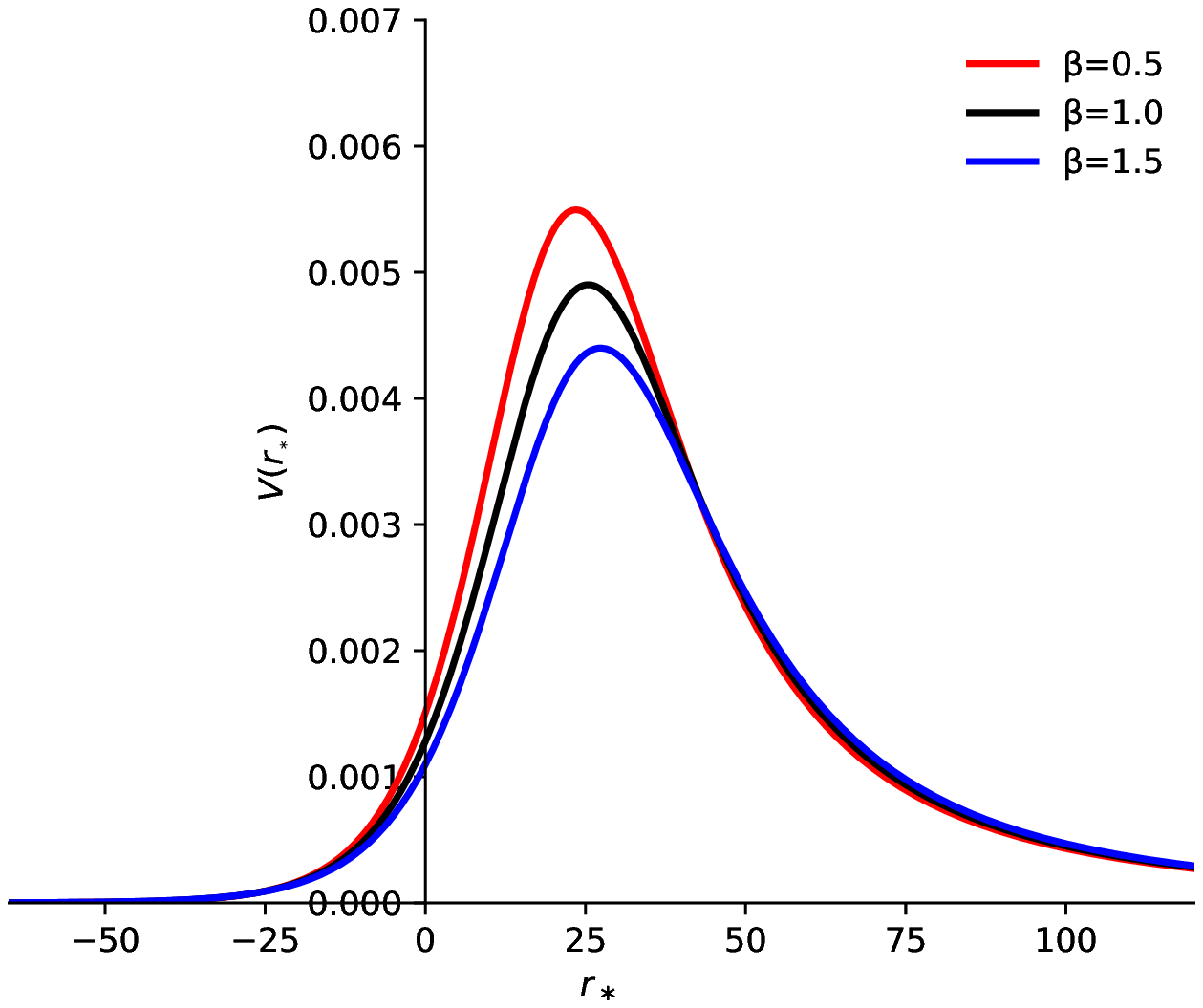}     
}
\subfigure[]{ 
\label{fig:b}     
\includegraphics[width=0.3\columnwidth]{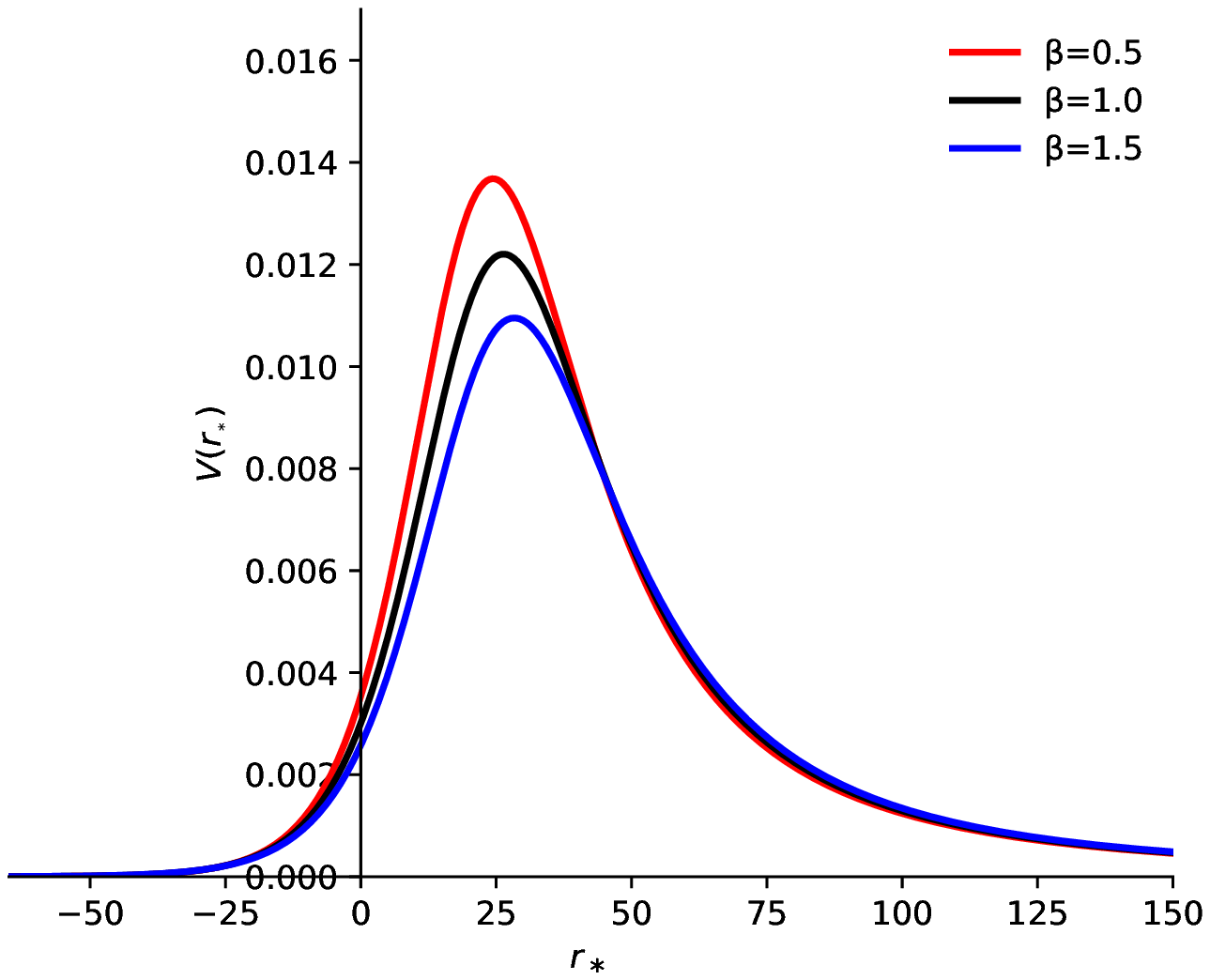}     
}
\caption{The effective potentials of the scalar field with the different $l$. (a) $l=0$ (b) $l=1$ (c) $l=2$.}     
\label{fig:1}
\end{figure*}

\begin{figure*}[t!]
\centering
\subfigure[]{ 
\label{fig:b}     
\includegraphics[width=0.3\columnwidth]{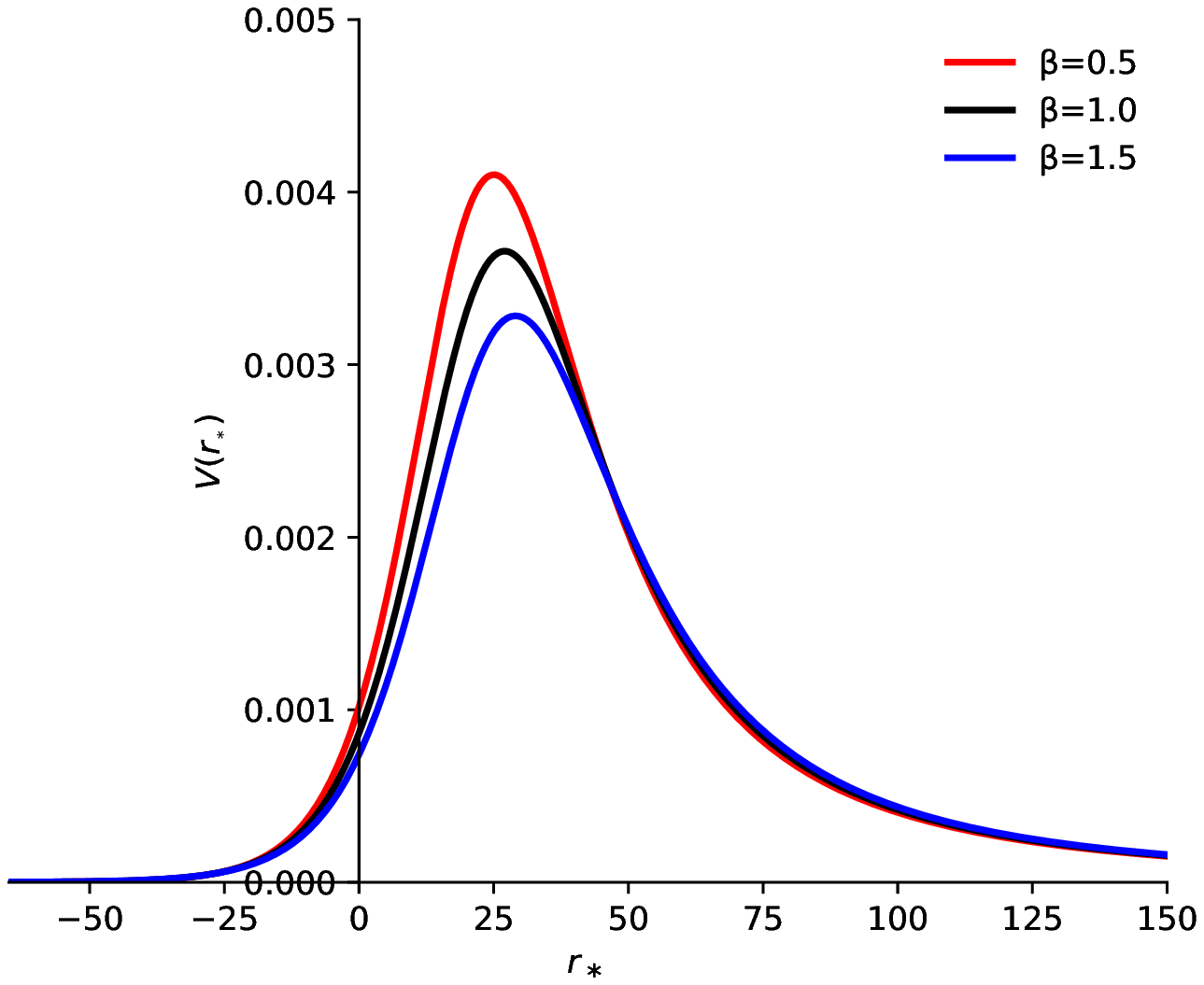}     
} 
\subfigure[]{ 
\label{fig:b}     
\includegraphics[width=0.3\columnwidth]{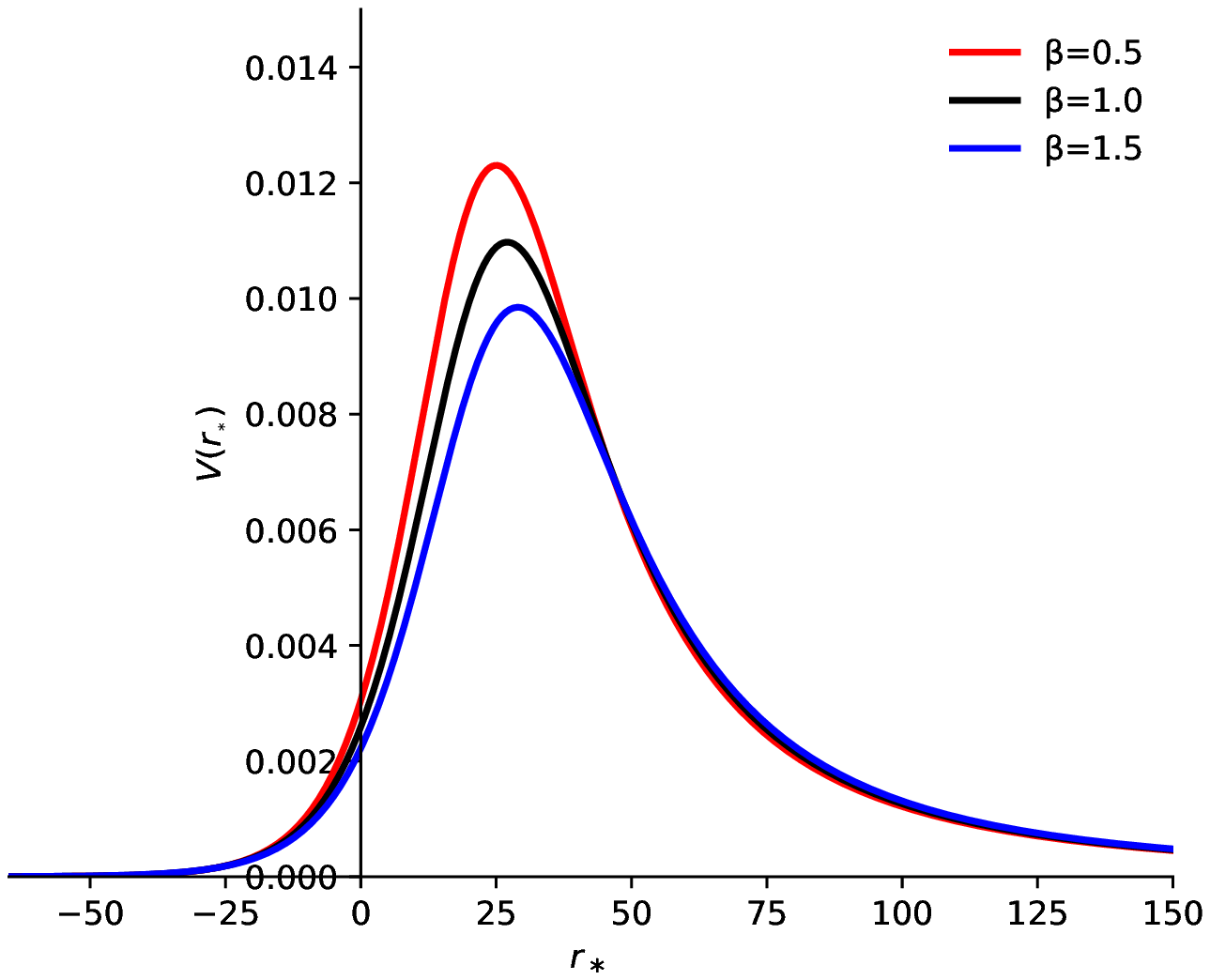}     
}
\subfigure[]{ 
\label{fig:b}     
\includegraphics[width=0.3\columnwidth]{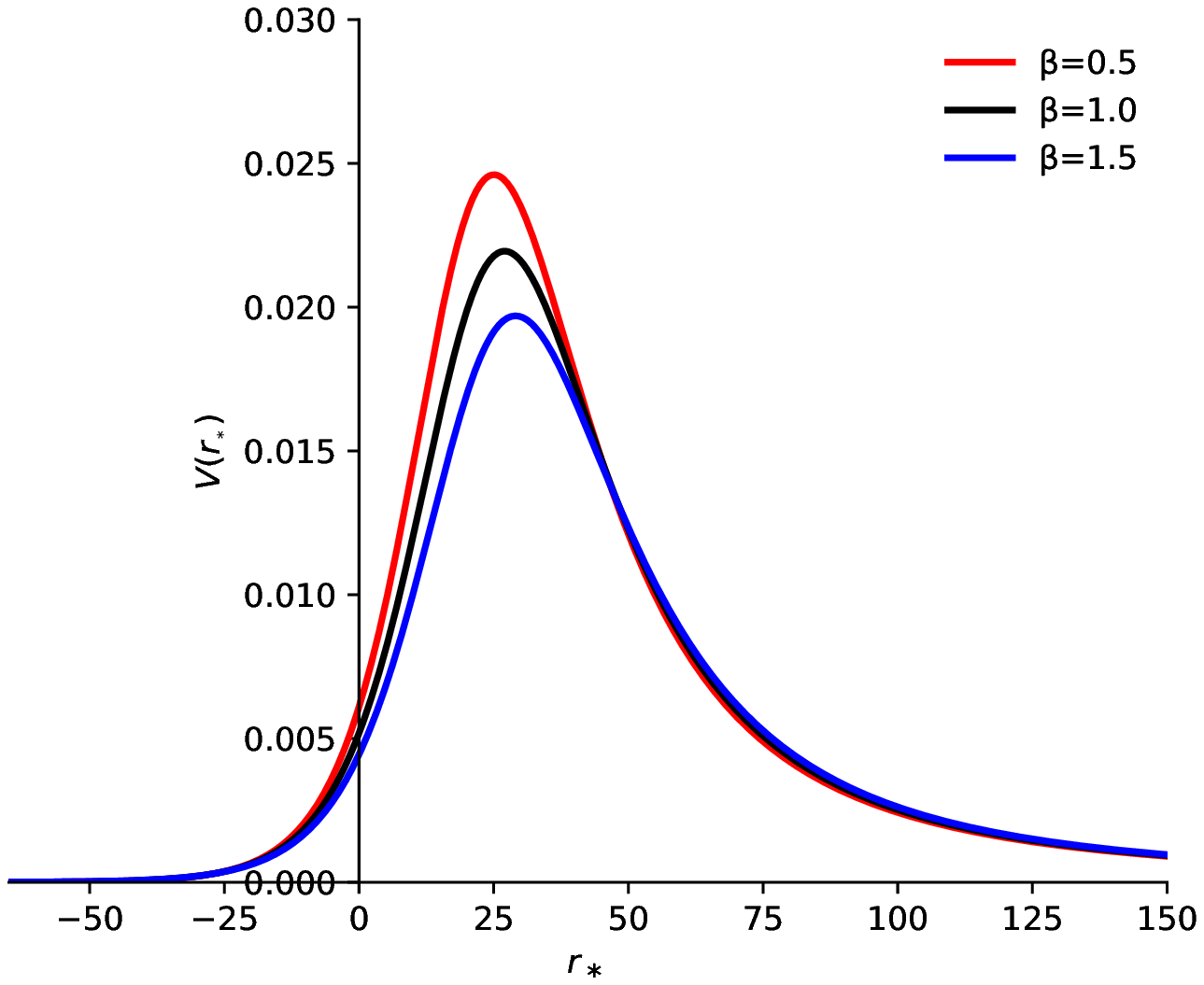}     
}
\caption{The effective potentials of the electromagnetic field with the different $l$. (a) $l=1$ (b) $l=2$ (c) $l=3$.}     
\label{fig:2}
\end{figure*}

\begin{figure*}[t!]
\centering
\subfigure[]{ 
\label{fig:b}     
\includegraphics[width=0.3\columnwidth]{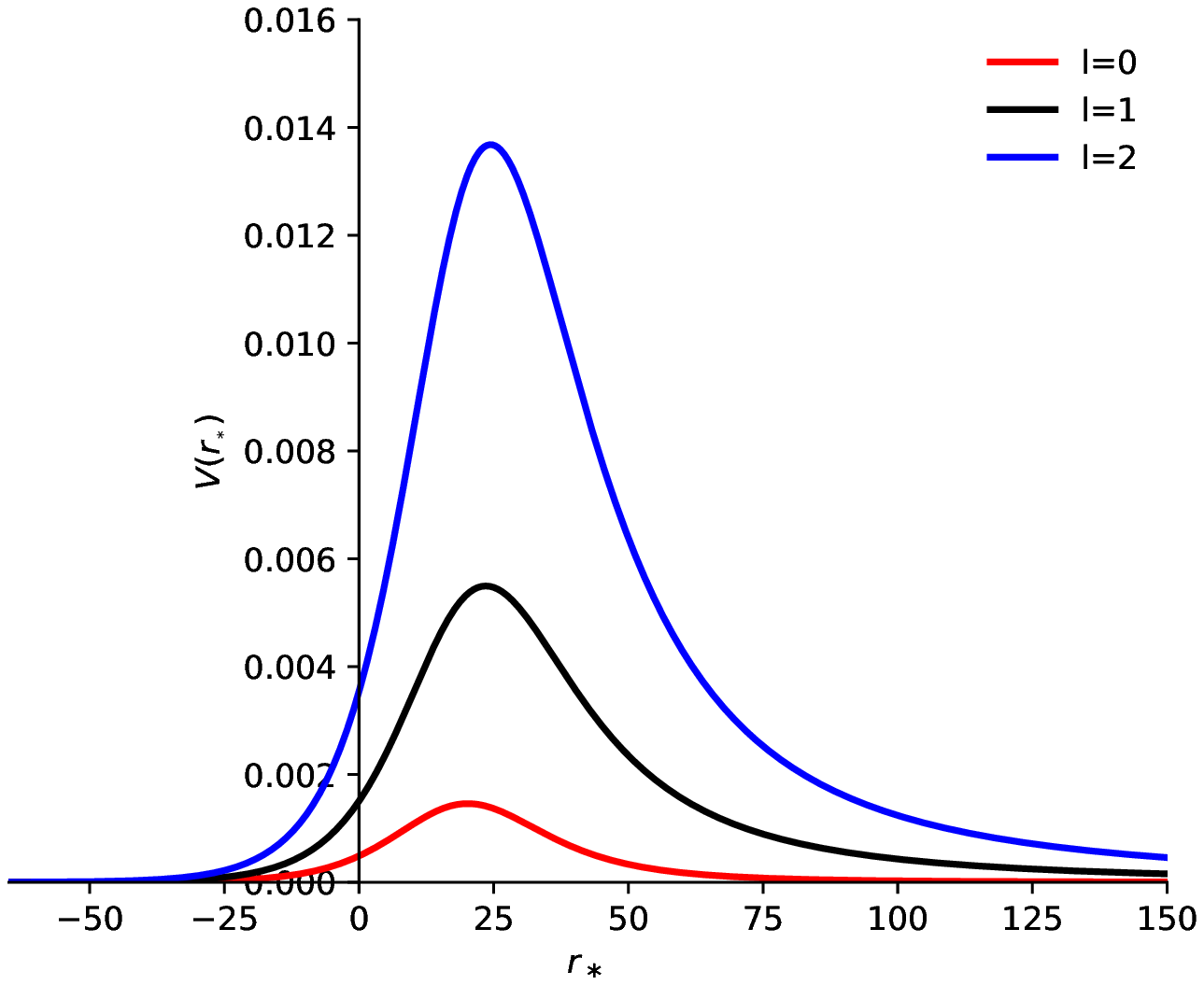}     
} 
\subfigure[]{ 
\label{fig:b}     
\includegraphics[width=0.3\columnwidth]{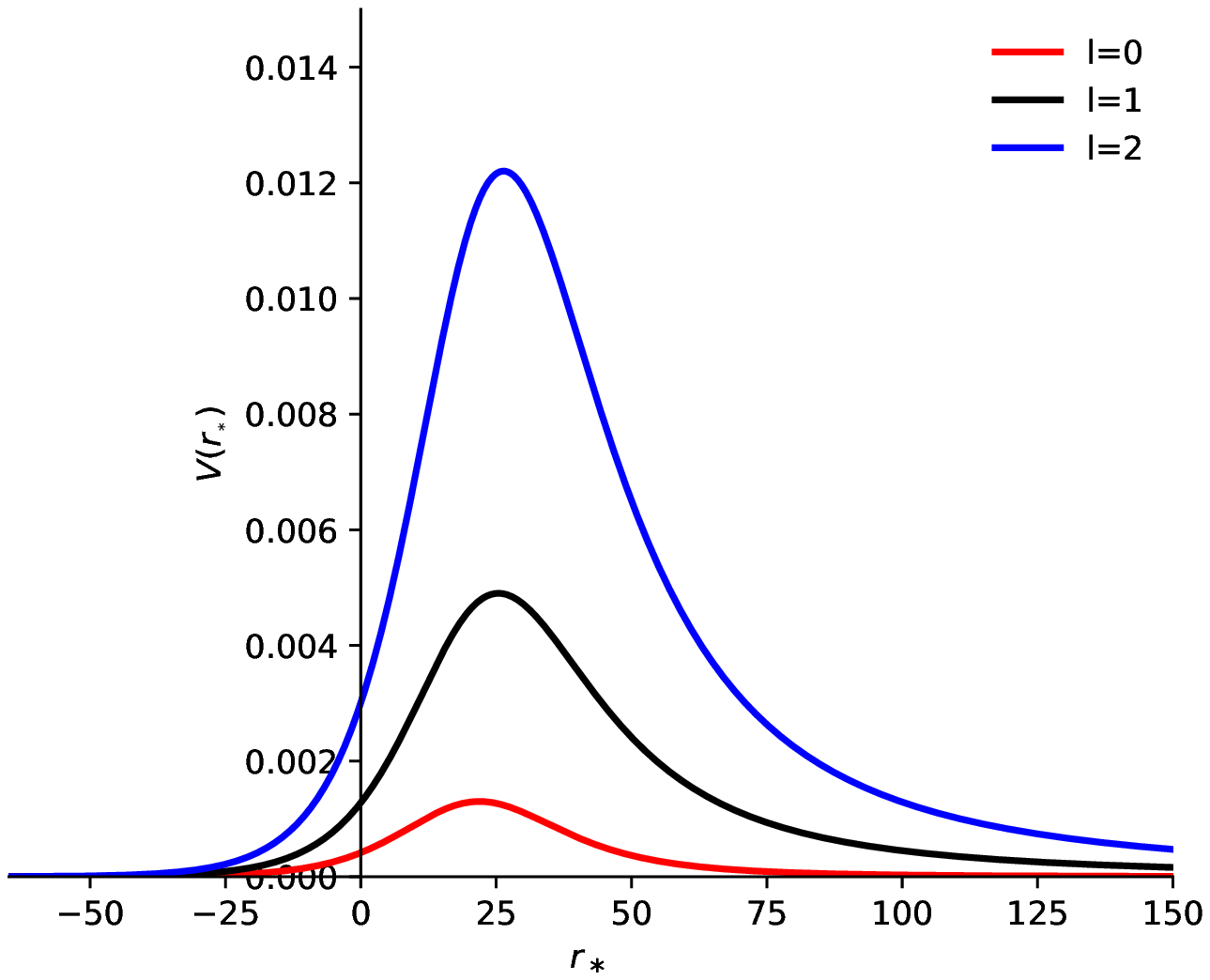}     
}
\subfigure[]{ 
\label{fig:b}     
\includegraphics[width=0.3\columnwidth]{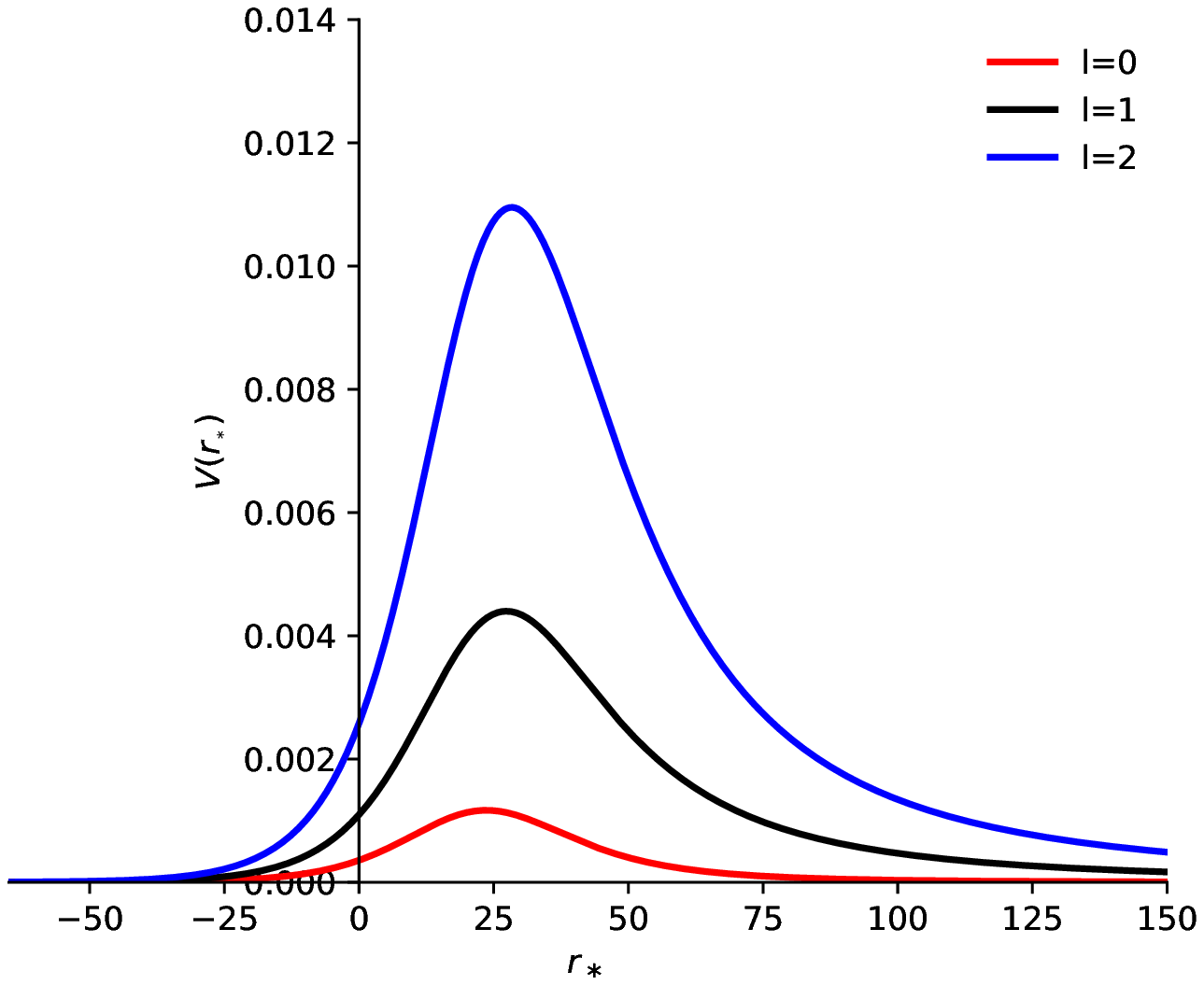}     
}
\caption{The effective potentials of the scalar field with the different $\beta$. (a) $\beta=0.5$ (b) $\beta=1.0$ (c) $\beta=1.5$.}     
\label{fig:3}
\end{figure*}

\begin{figure*}[t!]
\centering
\subfigure[]{ 
\label{fig:b}     
\includegraphics[width=0.3\columnwidth]{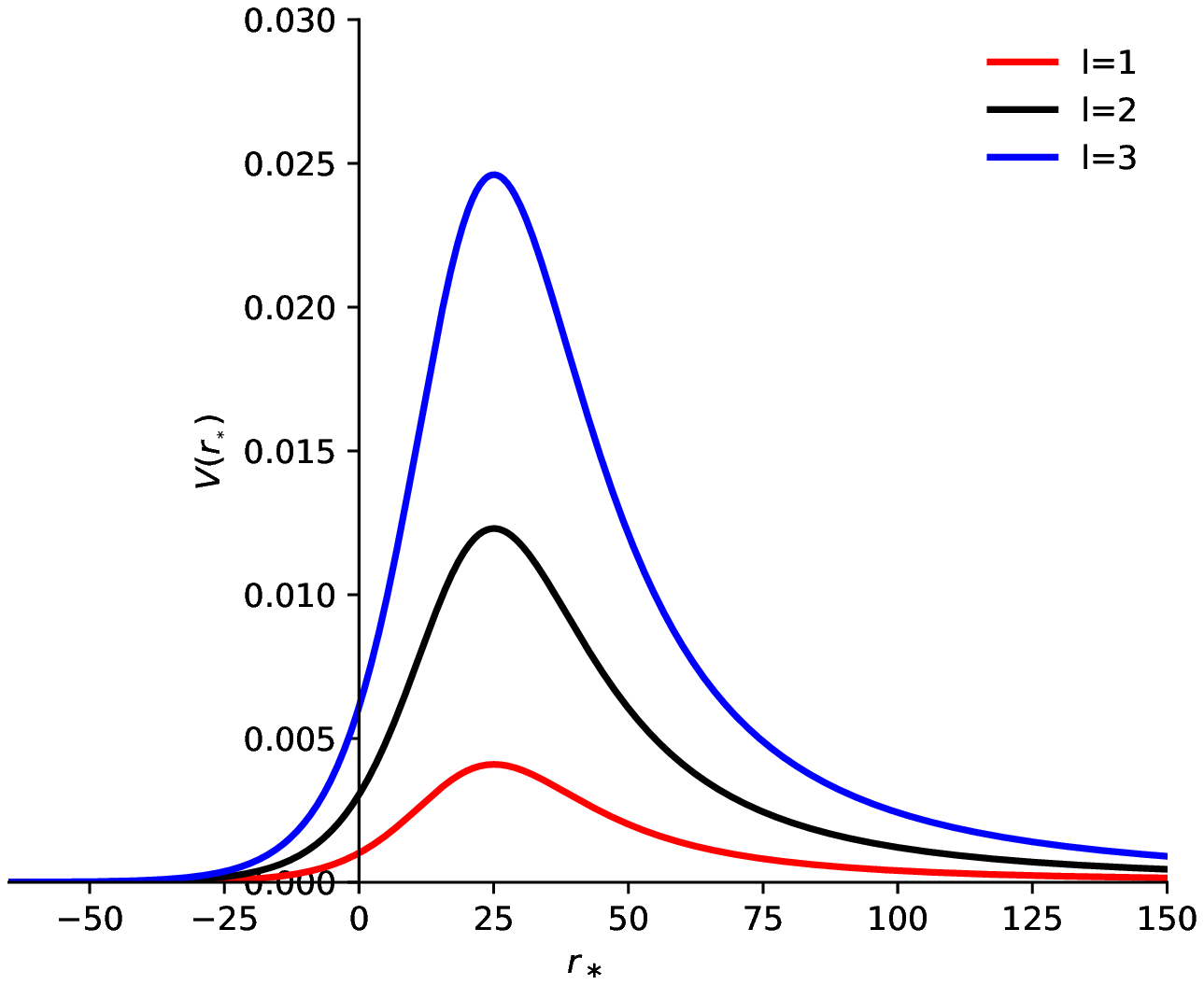}     
} 
\subfigure[]{ 
\label{fig:b}     
\includegraphics[width=0.3\columnwidth]{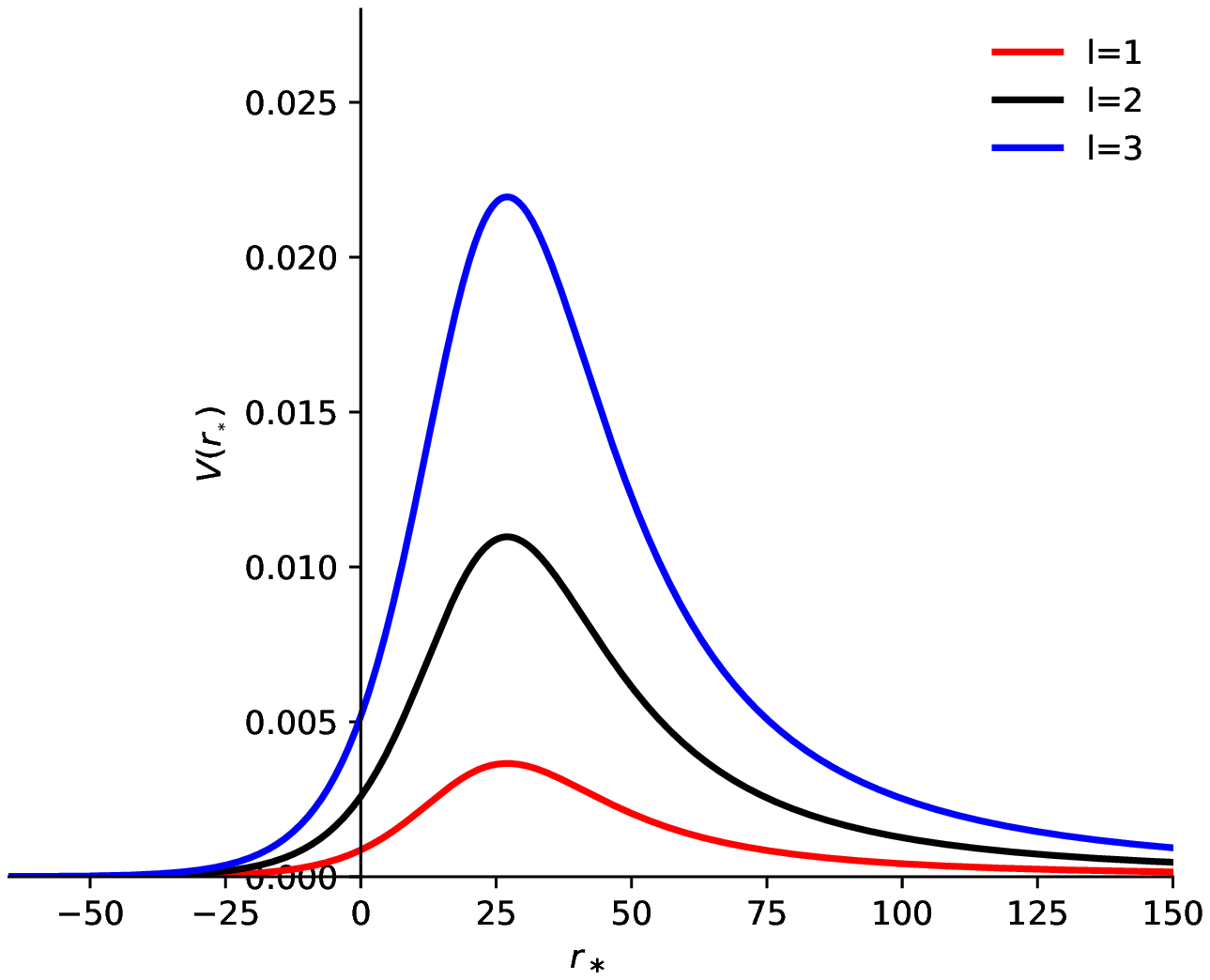}     
}
\subfigure[]{ 
\label{fig:b}     
\includegraphics[width=0.3\columnwidth]{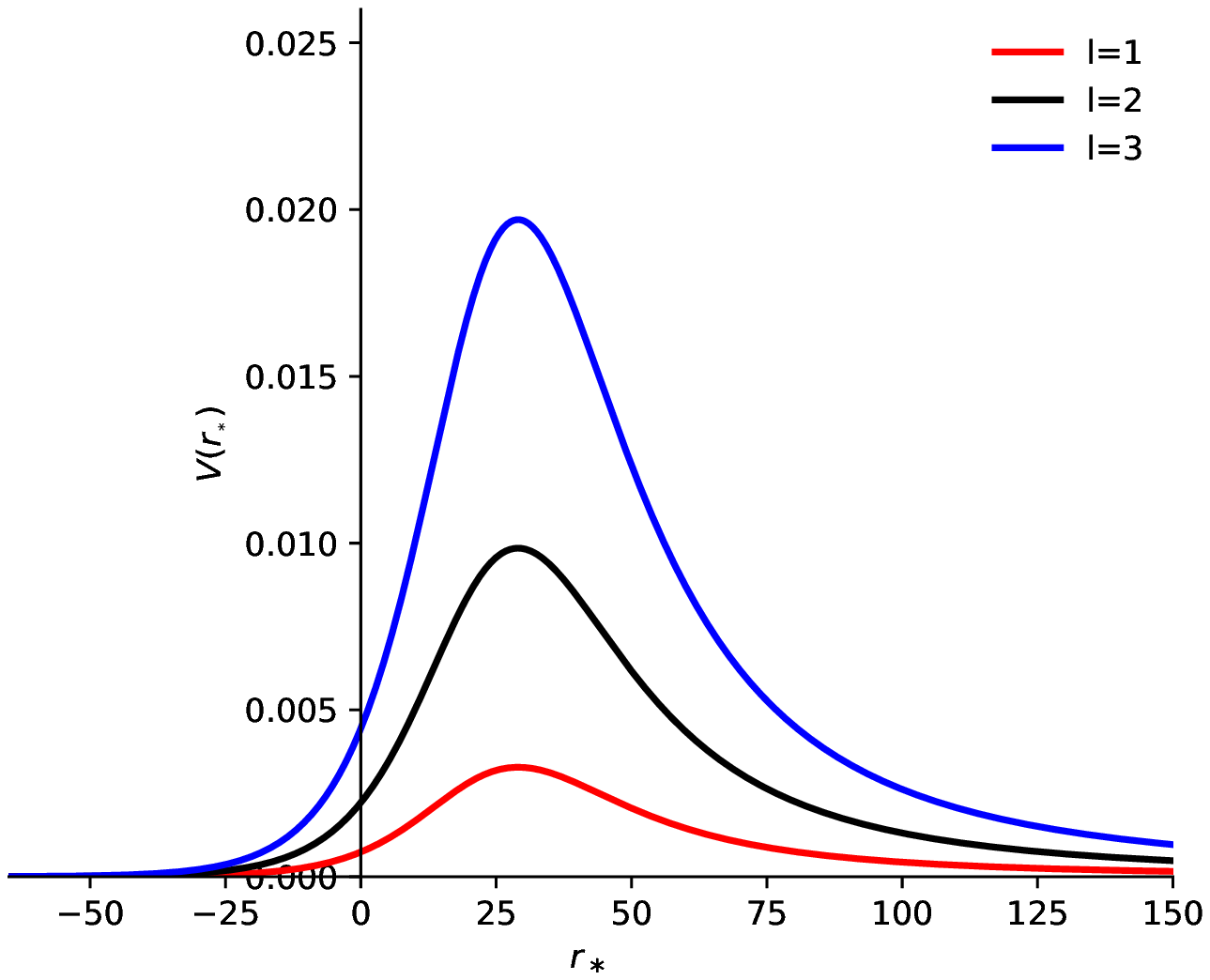}     
}
\caption{The effective potentials of the electromagnetic field with the different $\beta$. (a) $\beta=0.5$ (b) $\beta=1.0$ (c) $\beta=1.5$.}     
\label{fig:4}
\end{figure*}

\begin{figure*}[t!]
\centering
\subfigure[]{ 
\label{fig:b}     
\includegraphics[width=0.4\columnwidth]{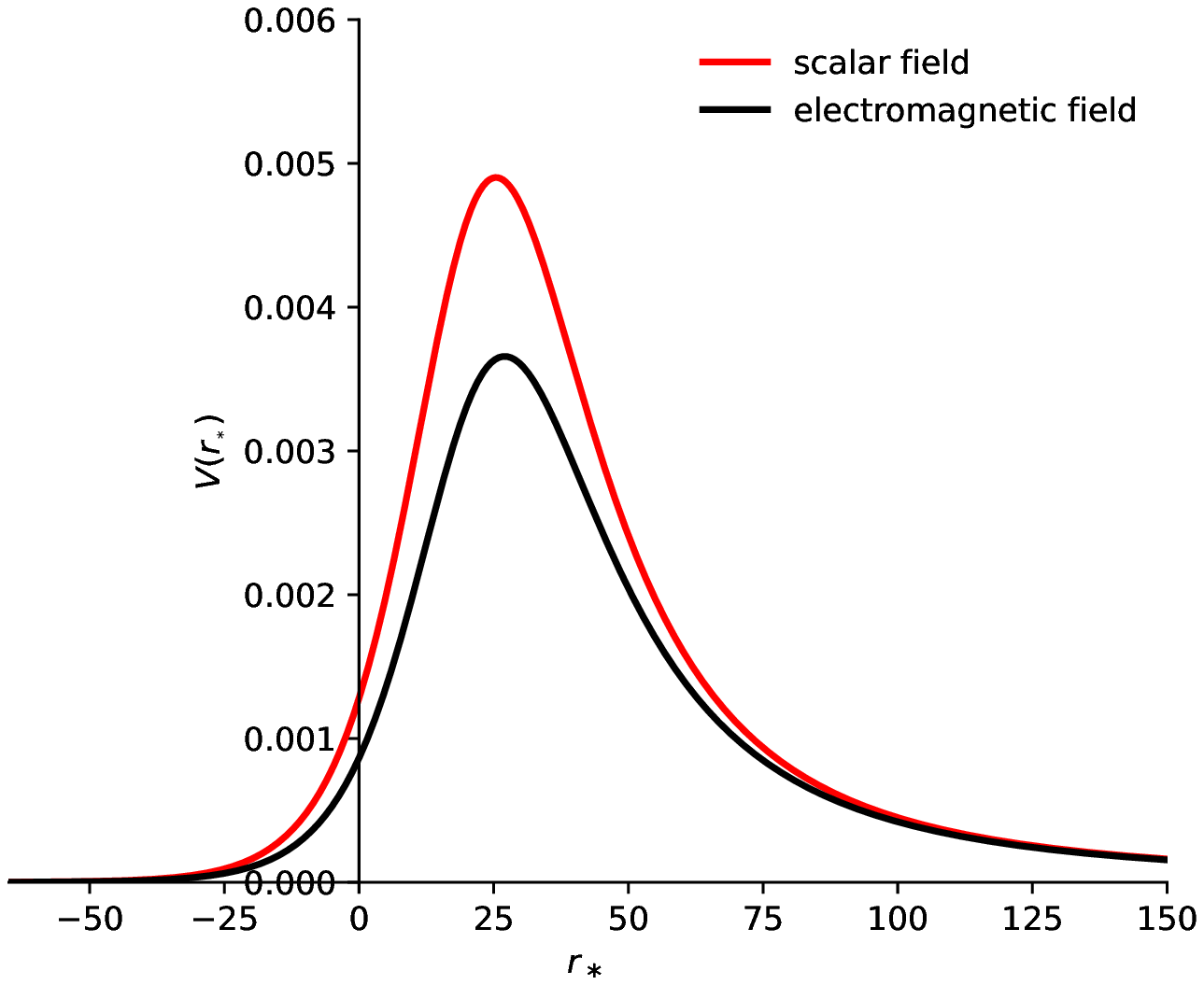}     
} 
\subfigure[]{ 
\label{fig:b}     
\includegraphics[width=0.4\columnwidth]{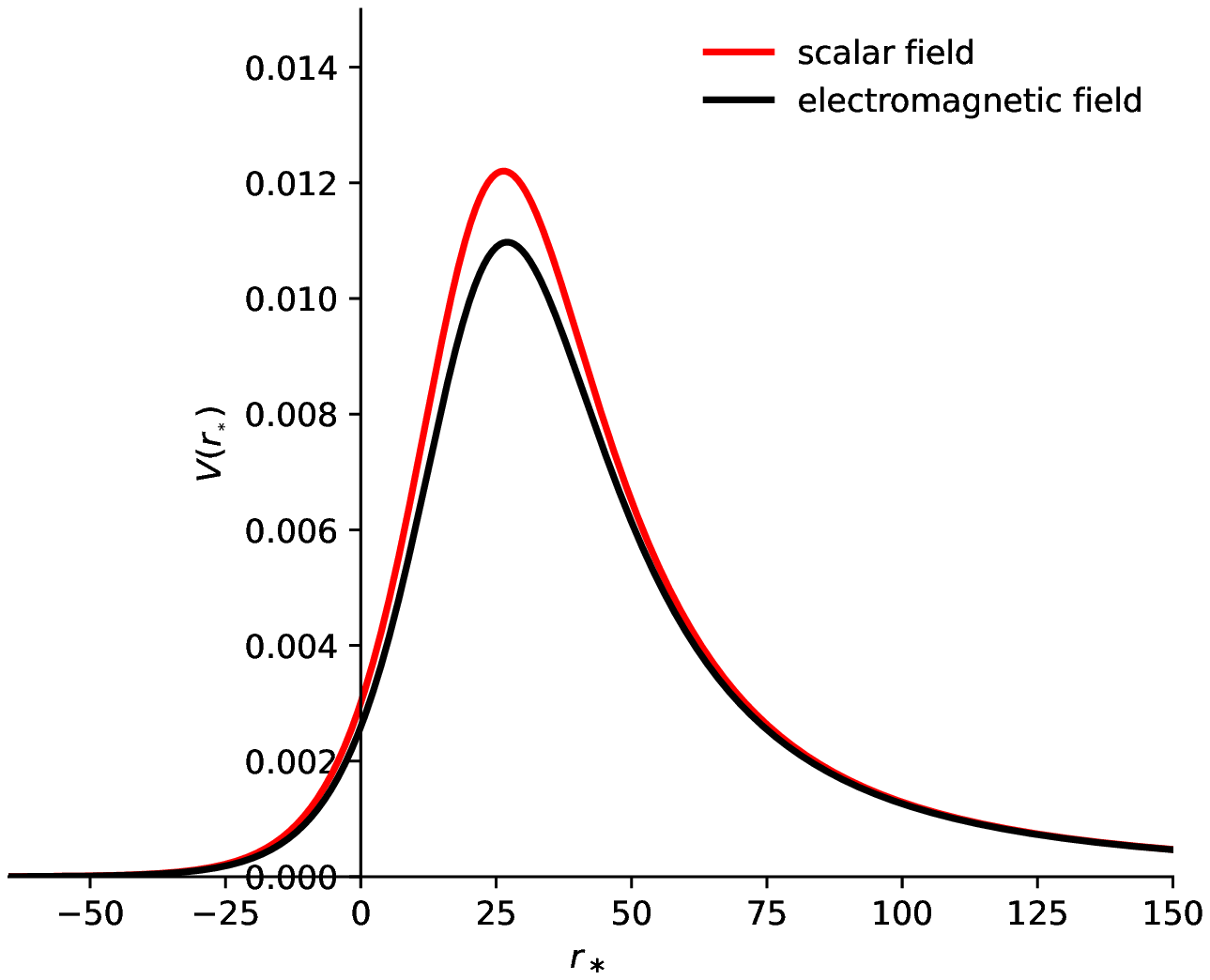}     
}

\caption{The effective potentials of the scalar and electromagnetic field. (a) $\beta=1, l=1$ (b) $\beta=1,l=2$.}     
\label{fig:4.1}
\end{figure*}

\indent Figs.\ref{fig:1}-\ref{fig:4} show the effective potential for various parameters $l$ and $\beta$. Figs.\ref{fig:1}-\ref{fig:2} show that under the scalar and electromagnetic fields, when $l$ is constant, the effective potential decreases with the increase of $\beta$, decays at infinity, and finally disappears, and then the black hole will return to the equilibrium state. Figs.\ref{fig:3}-\ref{fig:4} show that when $\beta$ is constant, the effective potential increases as $l$ increases, and when $r_{*}$ tends to infinity the effective potential approaches  $0$. Figs.\ref{fig:4.1} show that when the parameters $l$ and $\beta$ are fixed, the effective potential of the scalar field is more than the effective potential of the electromagnetic field. In \cite{ppp1,ppp2}, the QNM and effective potential of black holes in Einstein-Ethernet theory and bumblebee gravity are studied with great research significance, where the effective potential increases with $l$. The effective potential of the scalar field is larger than the electromagnetic field when $l$ is the same, and our conclusions are consistent with the results of this study.

\subsection{The WKB method}
\indent The WKB method was firstly introduced by Schutz and Will \cite{1} and then further explained by Konoplya \cite{2,3,4}. We use the WKB method to study frequencies, the main idea of which is to expand the WKB series in the asymptotic region and to match the solution of this expanding at infinity with the Taylor unfolding around the effective potential peak. In this session, we study the QNM frequencies by means of the sixth-order WKB formula of the shape given below:
\begin{equation}
\frac{\text{i}(\omega ^{2}-V_{0})}{\sqrt{-2V_{0}^{''}}}-\sum ^{6}_{\text{i}=2}\Omega_\text{i}=n+\frac{1}{2},  
\label{equ10}                                                   
\end{equation}
where $V_{0}$ is the maximal value of the effective potential at $r_\text{0}$, $V_\text{0}^{''}$ is the second order of the derivatives of the tortoise coordinates, $\Omega_\text{i}$ are the correction terms. 
\subsection{The time domain method}
\indent The time-domain integration method proposed by Gundlach, Price and Pullin \cite{24,25} to study the process of change of dynamical equations. We use this method to calculate the variation of frequency with time in the background spacetime of the black hole. By introducing the light cone coordinates for numerical computation
\begin{equation}
\begin{array}{l}
u=t-r_*, \\
v=t+r_*,
\label{equ14}
\end{array}
\end{equation}
where $u$ and $v$ are integration constants, $r_*$ is tortoise coordinate. The wave function equation is being given as
\begin{eqnarray}
-4\frac{\partial ^{2}\psi (\mu ,\nu )}{\partial \mu \partial \nu }=V(\mu ,\nu)\psi (\mu ,\nu ).
\label{equ15}
\end{eqnarray}
\indent The effective potential is influenced by the light cone variable (\ref{equ14}). The difference step of the discretization is shown in \cite{25}, this discretization scheme is given by

\begin{eqnarray}  
\Psi(N)=\Psi(W)+\Psi(E)-\Psi(S)-h ^2\frac{V(W)\Psi(W)+V(E)\Psi(E)}{8}+O(h ^4).     
\label{equ16}                                             
\end{eqnarray}

\indent Where we use the following specified points $N=(u+h ,v+h)$, $W=(u+h,v)$, $E=(u,v +h)$ and $S = (u,v)$, where $\nu=\nu _{0}$, $\mu=\mu _{0}$ are the initial datas. In order to contain the accuracy of the frequency analysis, we can control the parameters of the mathematical combination, adjust the differential grid and accuracy \cite{26}.

\begin{table*}[t!]
\setlength{\abovecaptionskip}{0.5cm}
\setlength{\belowcaptionskip}{0.2cm}
\caption{The QNM frequencies of scalar field.}
\setlength{\tabcolsep}{3mm}{
\begin{tabular}{ccccc}
\hline \hline
\multicolumn{5}{c}{WKB method}  \\  \hline
$\beta$& $l=0$                  & $l=1$                  & $l=2$                     & $l=3$   \\ \hline
$0.25$ & 0.026779 - 0.024441i & 0.071008 - 0.023699i & 0.117247 - 0.023458i & 0.163725 - 0.023394i \\ 
$0.50$ & 0.025991 - 0.023722i & 0.068919 - 0.023002i & 0.113798 - 0.022768i & 0.158910 - 0.022706i   \\ 
$0.75$ & 0.025248 - 0.023044i & 0.066950 - 0.022345i & 0.110547 - 0.022118i & 0.154369 - 0.022057i \\ 
$1.00$ & 0.024547 - 0.022404i & 0.065091 - 0.021724i & 0.107476 - 0.021503i & 0.150081 - 0.021444i \\
$1.25$ & 0.023885 - 0.021797i & 0.063331 - 0.021130i & 0.104571 - 0.020922i & 0.146025 - 0.020865i \\ 
$1.50$ & 0.023255 - 0.021224i & 0.061665 - 0.020581i & 0.101819 - 0.020371i & 0.142182 - 0.020315i\\

\hline \hline
\end{tabular}}
\vspace{0.3cm}
\label{tab:1}
\end{table*}

\begin{table*}[t!]
\setlength{\abovecaptionskip}{0.5cm}
\setlength{\belowcaptionskip}{0.2cm}
\caption{The QNM frequencies of electromagnetic field.}
\setlength{\tabcolsep}{3mm}{
\begin{tabular}{ccccc}
\hline \hline
\multicolumn{5}{c}{WKB method}  \\  \hline
$\beta$& $l=1$                 & $l=2$                  & $l=3$                  & $l=4$   \\ \hline
$0.25$ & 0.0601676 -0.022457i & 0.110932 - 0.023033i & 0.159248 - 0.023179i & 0.206811 - 0.023238i\\
$0.50$ & 0.058398 - 0.021796i & 0.107669 - 0.022355i & 0.154564 - 0.022498i & 0.200728 - 0.022555i\\ 
$0.75$ & 0.0567295 -0.021174i & 0.104593 - 0.021716i & 0.150148 - 0.021855i & 0.194993 - 0.021910i\\
$1.00$ & 0.055153 - 0.020586i & 0.101687 - 0.021113i & 0.145977 - 0.021248i & 0.189577 - 0.021302i\\
$1.25$ & 0.053663 - 0.020029i & 0.098939 - 0.020541i & 0.142032 - 0.020674i & 0.184453 - 0.020726i\\
$1.50$ & 0.052250 - 0.019502i & 0.096335 - 0.020002i & 0.138294 - 0.020129i & 0.179599 - 0.020181i \\

\hline \hline
\end{tabular}}
\vspace{0.4cm}
\label{tab:2}
\end{table*}

\begin{figure*}[t!]
\centering
\subfigure[]{ 
\label{fig:b1}     
\includegraphics[width=0.3\columnwidth]{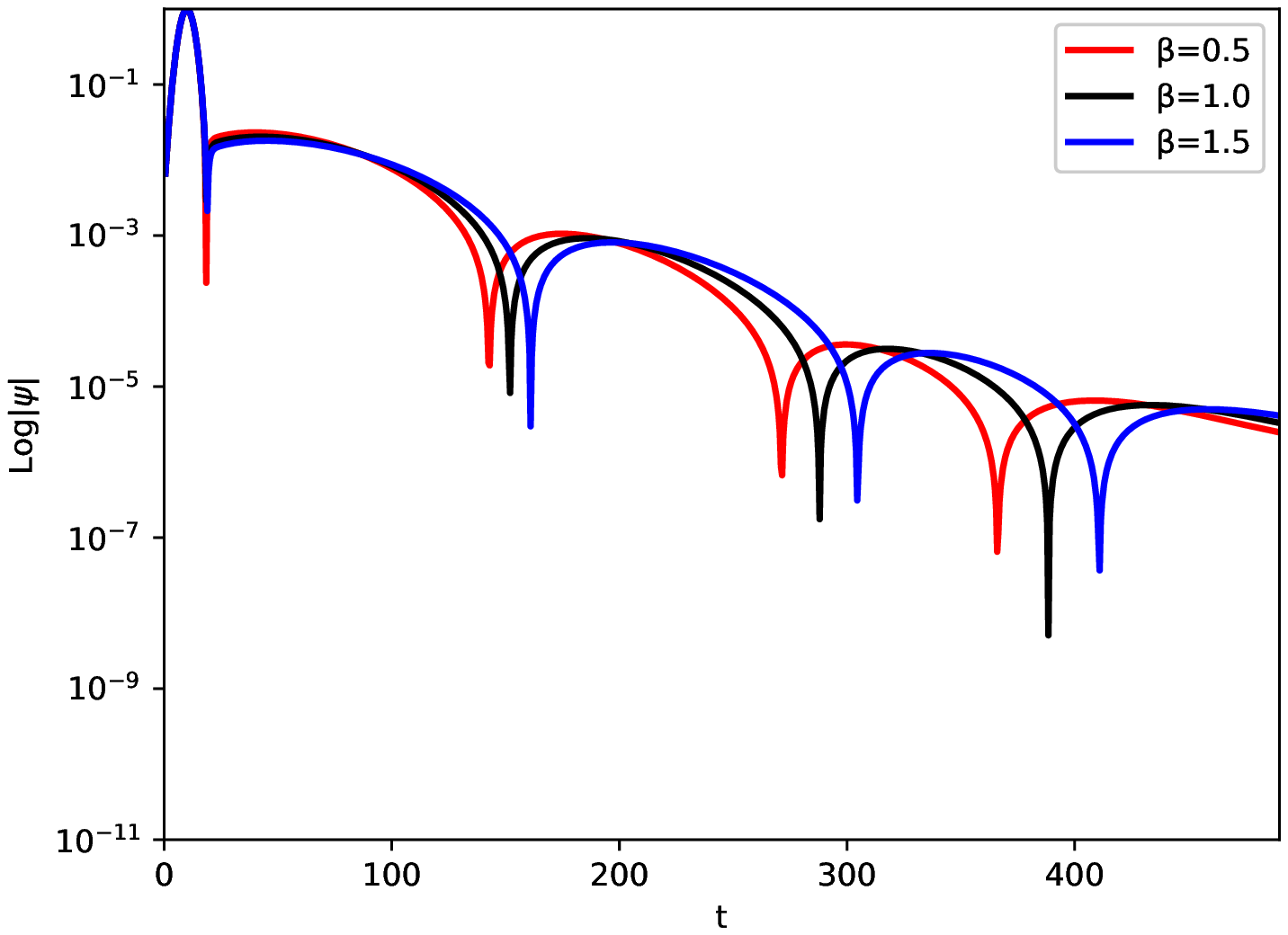}     
} 
\subfigure[]{ 
\label{fig:b2}     
\includegraphics[width=0.3\columnwidth]{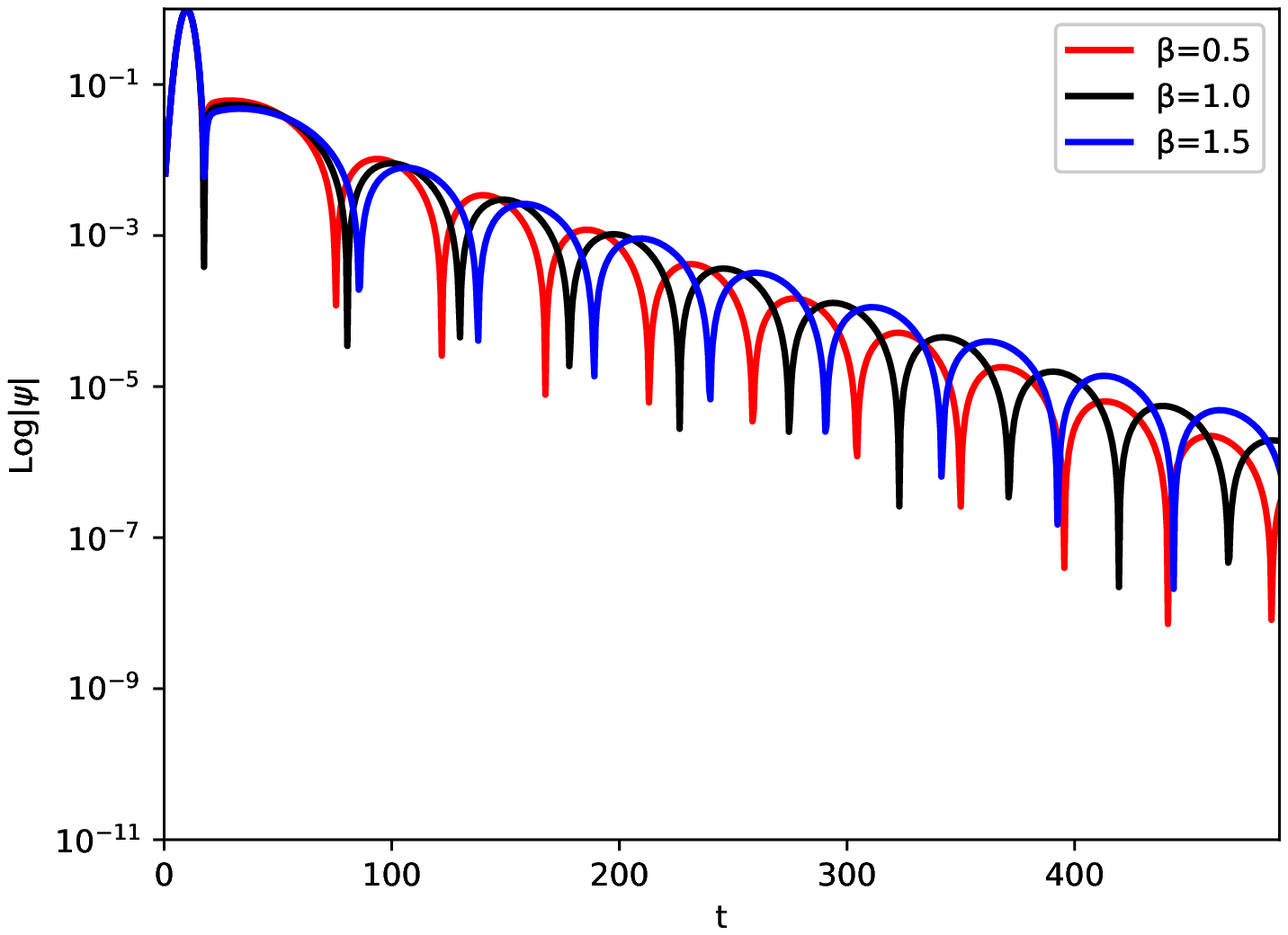}     
}
\subfigure[]{ 
\label{fig:b3}     
\includegraphics[width=0.3\columnwidth]{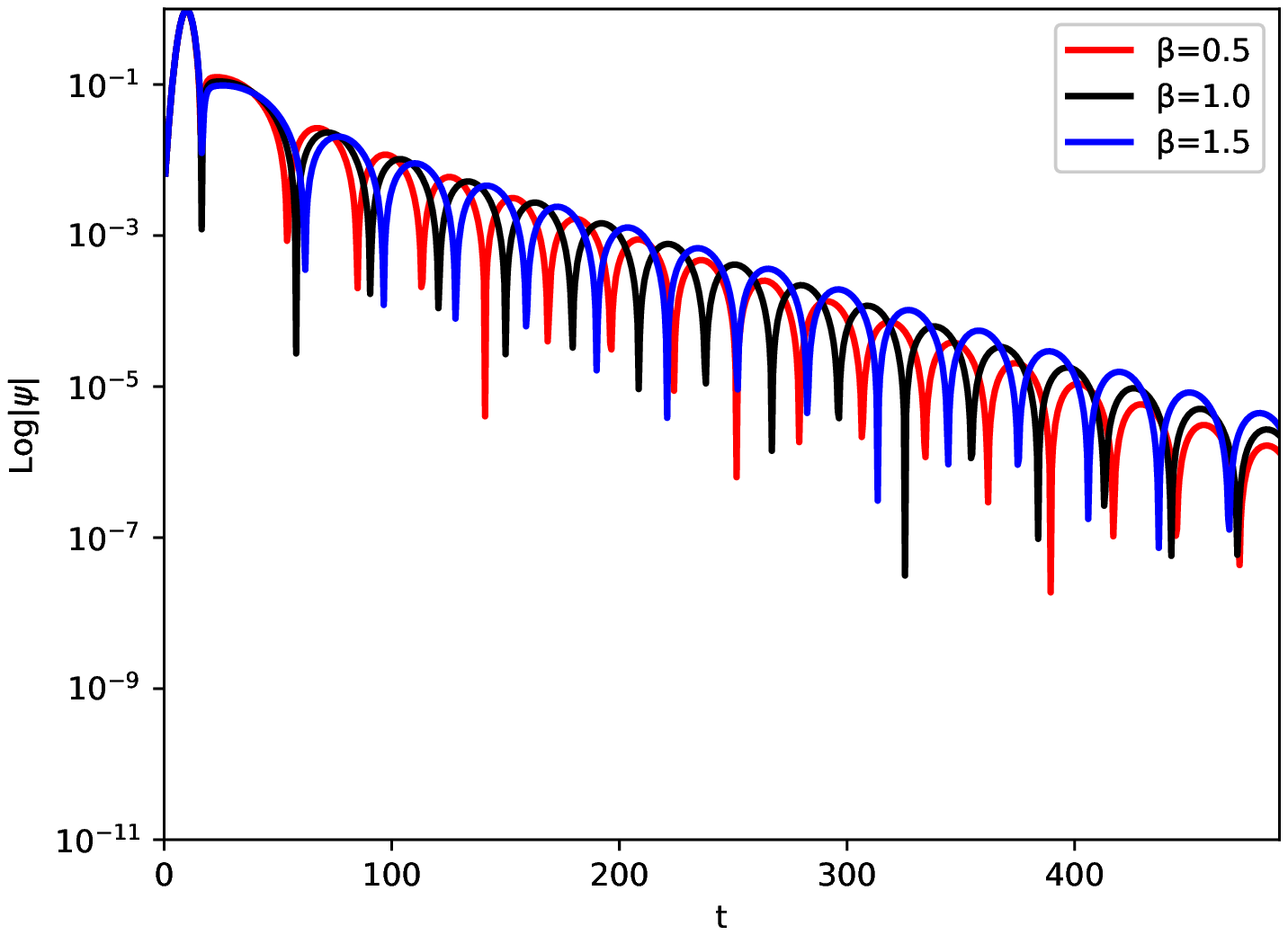}     
}
\caption{The dynamical evolutions of the scalar field with the different $l$. (a) $l=0$ (b) $l=1$ (c) $l=2$.}     
\label{fig:5}
\end{figure*}

\begin{figure*}[t!]
\centering
\subfigure[]{ 
\label{fig:b}     
\includegraphics[width=0.3\columnwidth]{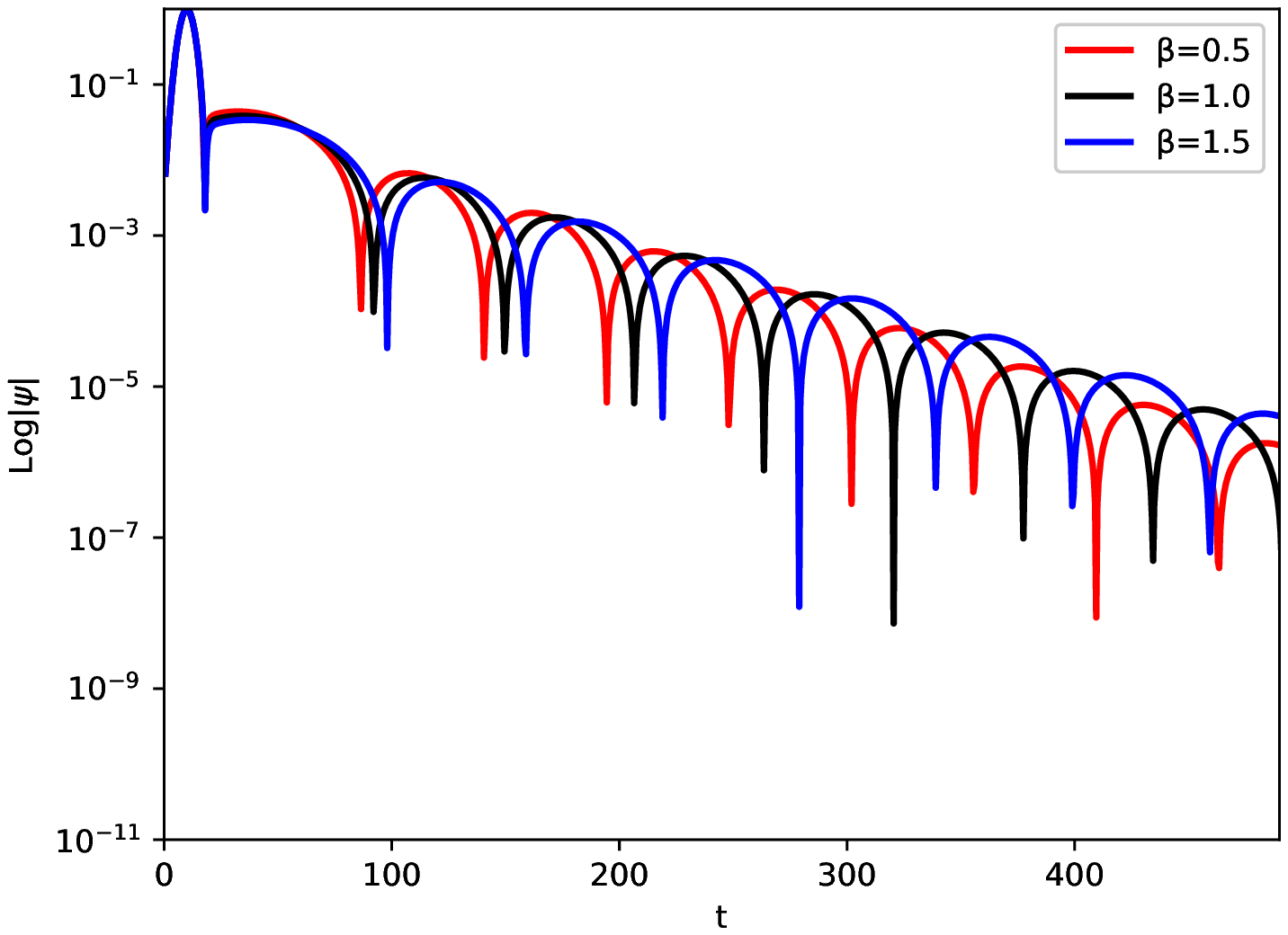}     
} 
\subfigure[]{ 
\label{fig:b}     
\includegraphics[width=0.3\columnwidth]{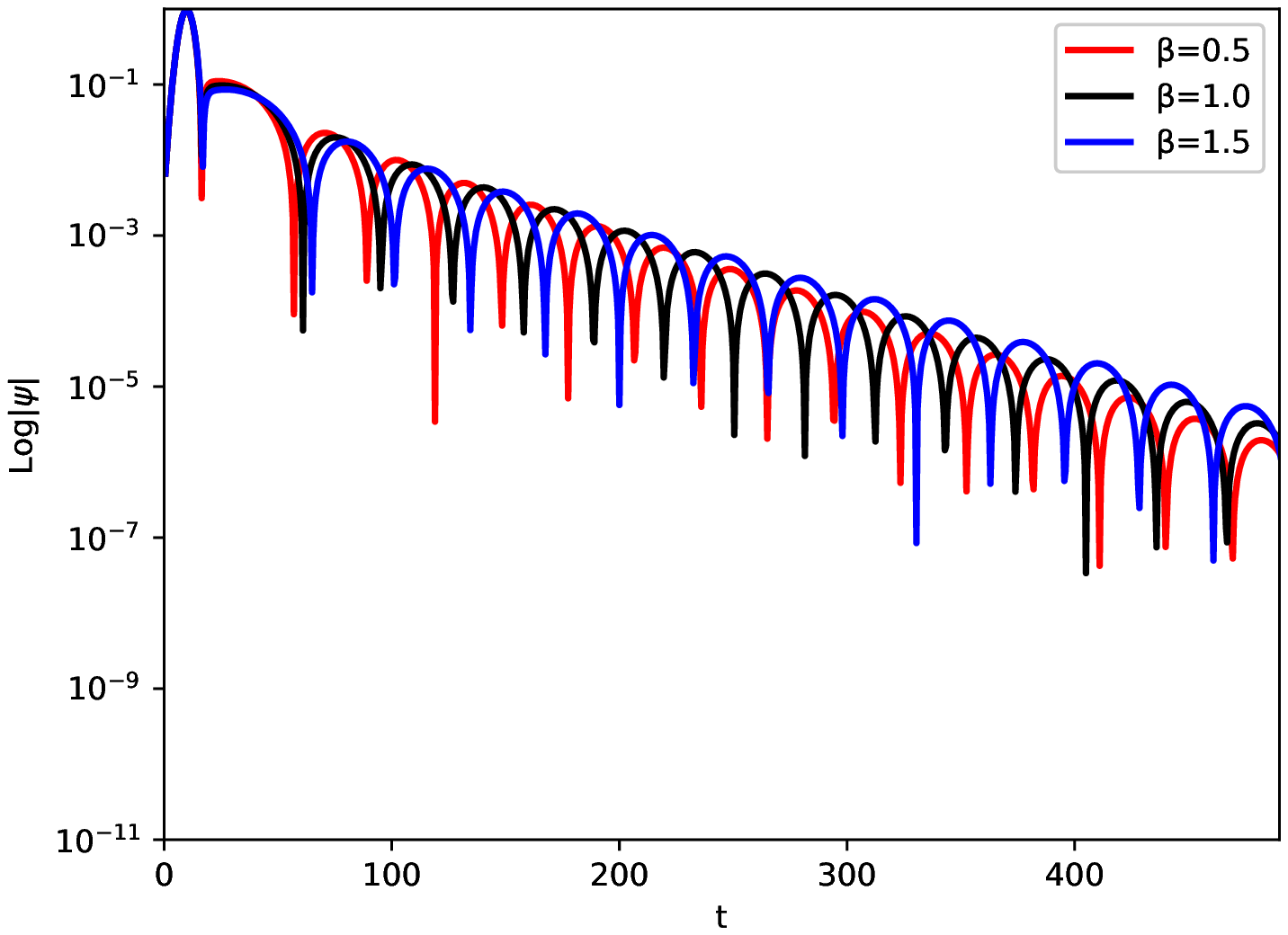}     
}
\subfigure[]{ 
\label{fig:b}     
\includegraphics[width=0.3\columnwidth]{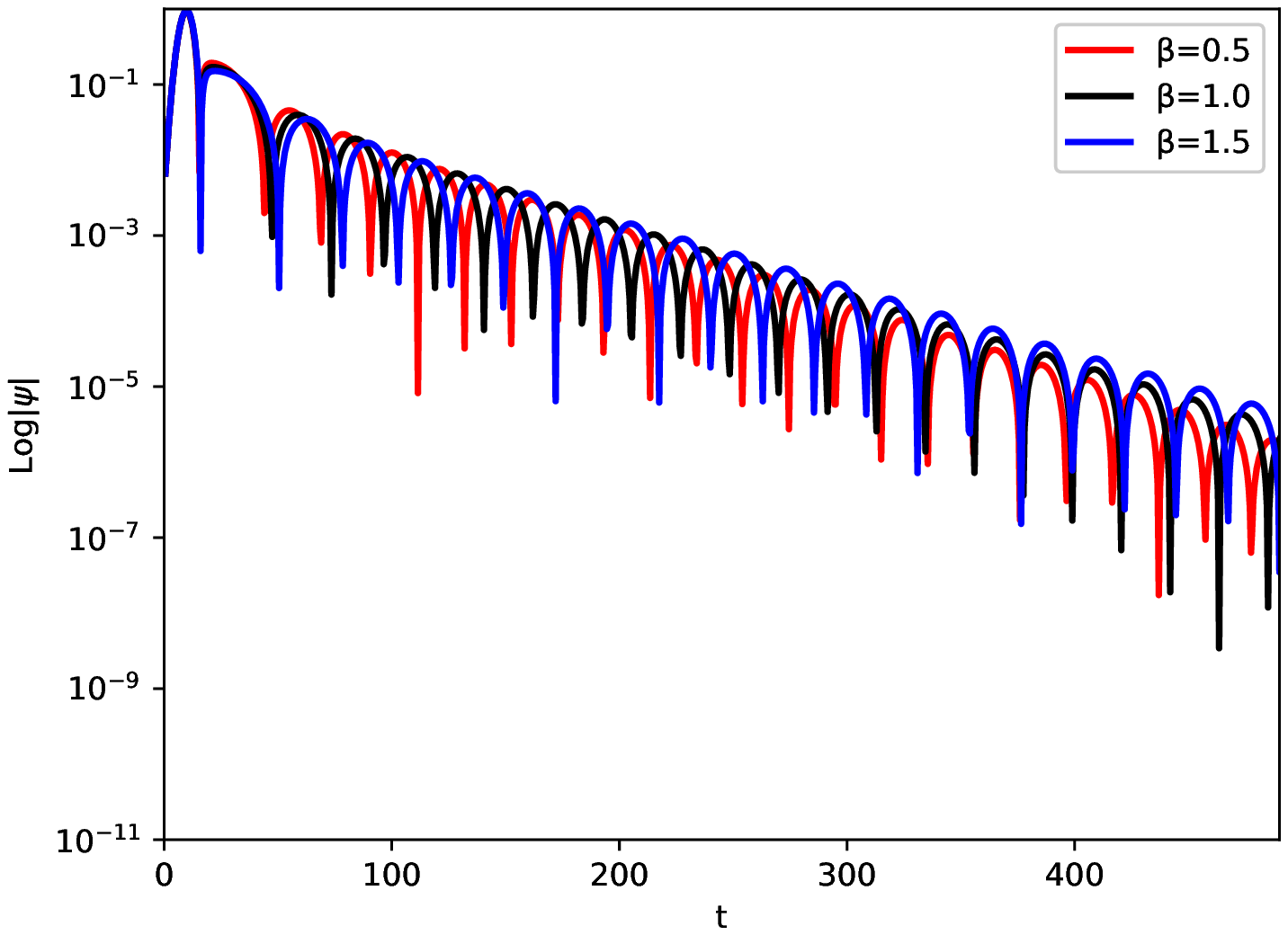}     
}
\caption{The dynamical evolutions of the electromagnetic field with the different $l$. (a) $l=1$ (b) $l=2$ (c) $l=3$.}     
\label{fig:6}
\end{figure*}

\begin{figure*}[t!]
\centering
\subfigure[]{ 
\label{fig:b}     
\includegraphics[width=0.3\columnwidth]{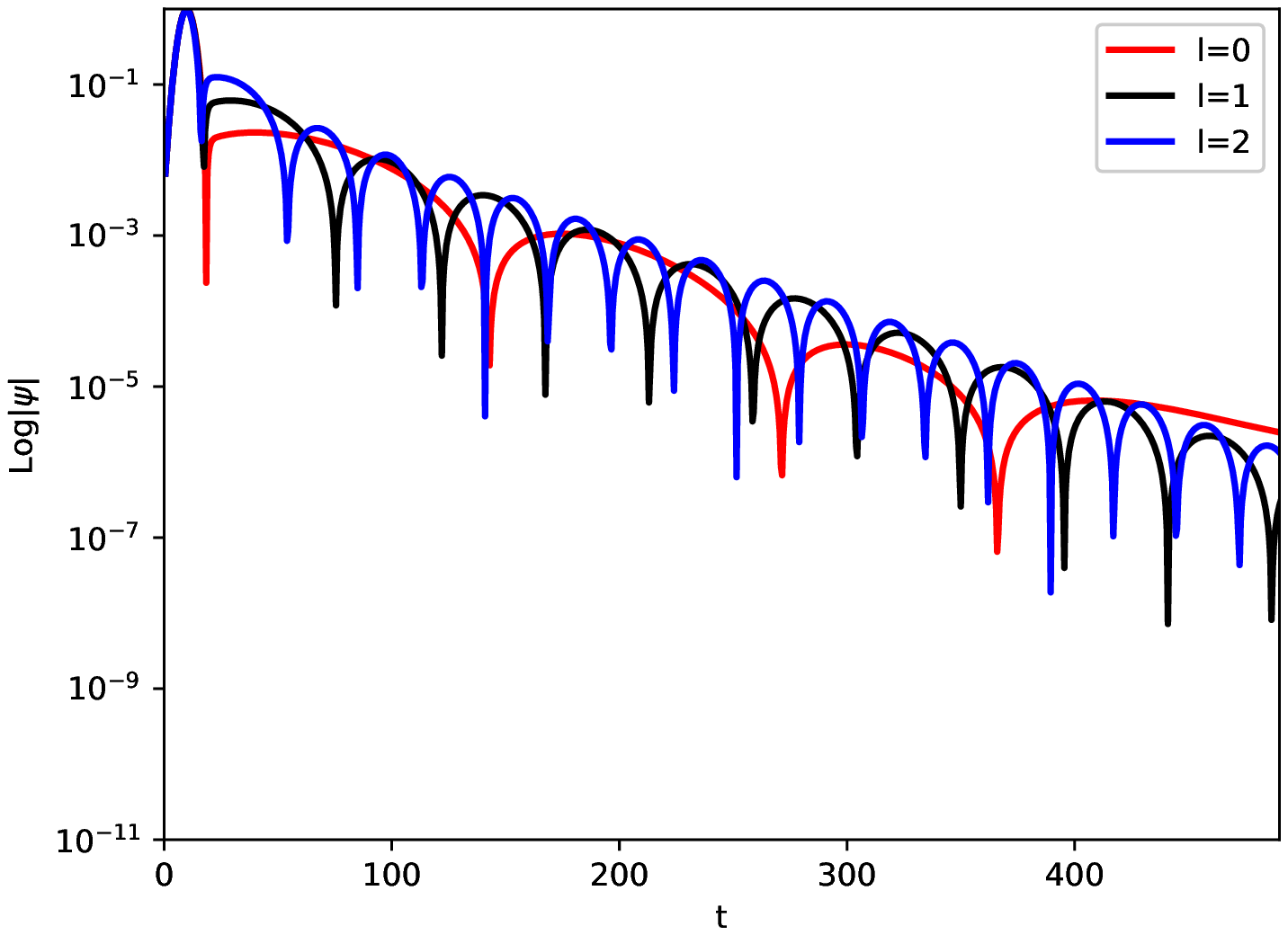}     
} 
\subfigure[]{ 
\label{fig:b}     
\includegraphics[width=0.3\columnwidth]{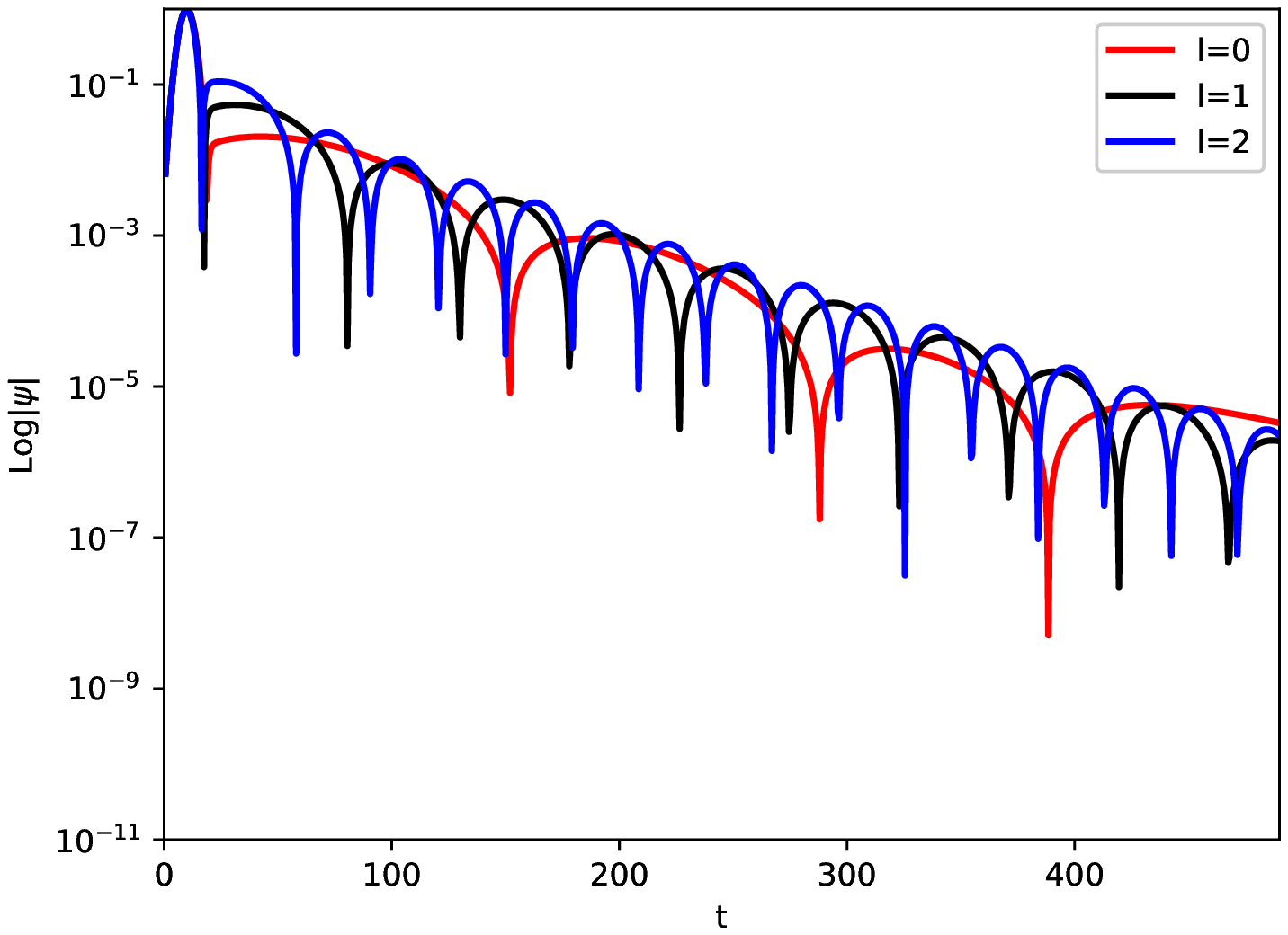}     
}
\subfigure[]{ 
\label{fig:b}     
\includegraphics[width=0.3\columnwidth]{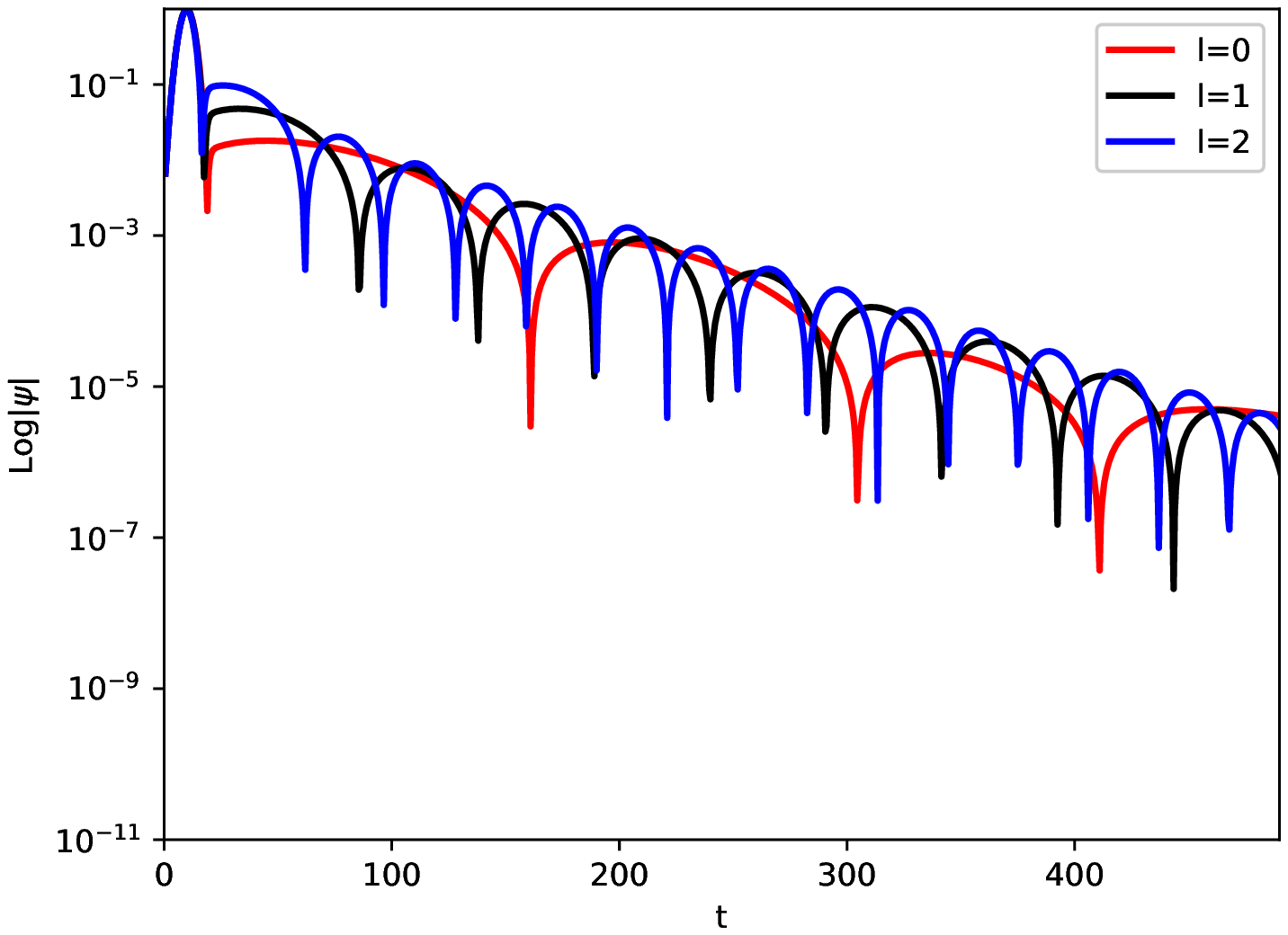}     
}
\caption{The dynamical evolutions of the scalar field with the different $\beta$. (a) $\beta=0.5$ (b) $\beta=1.0$ (c) $\beta=1.5$.}     
\label{fig:7}
\end{figure*}

\begin{figure*}[t!]
\centering
\subfigure[]{ 
\label{fig:b}     
\includegraphics[width=0.3\columnwidth]{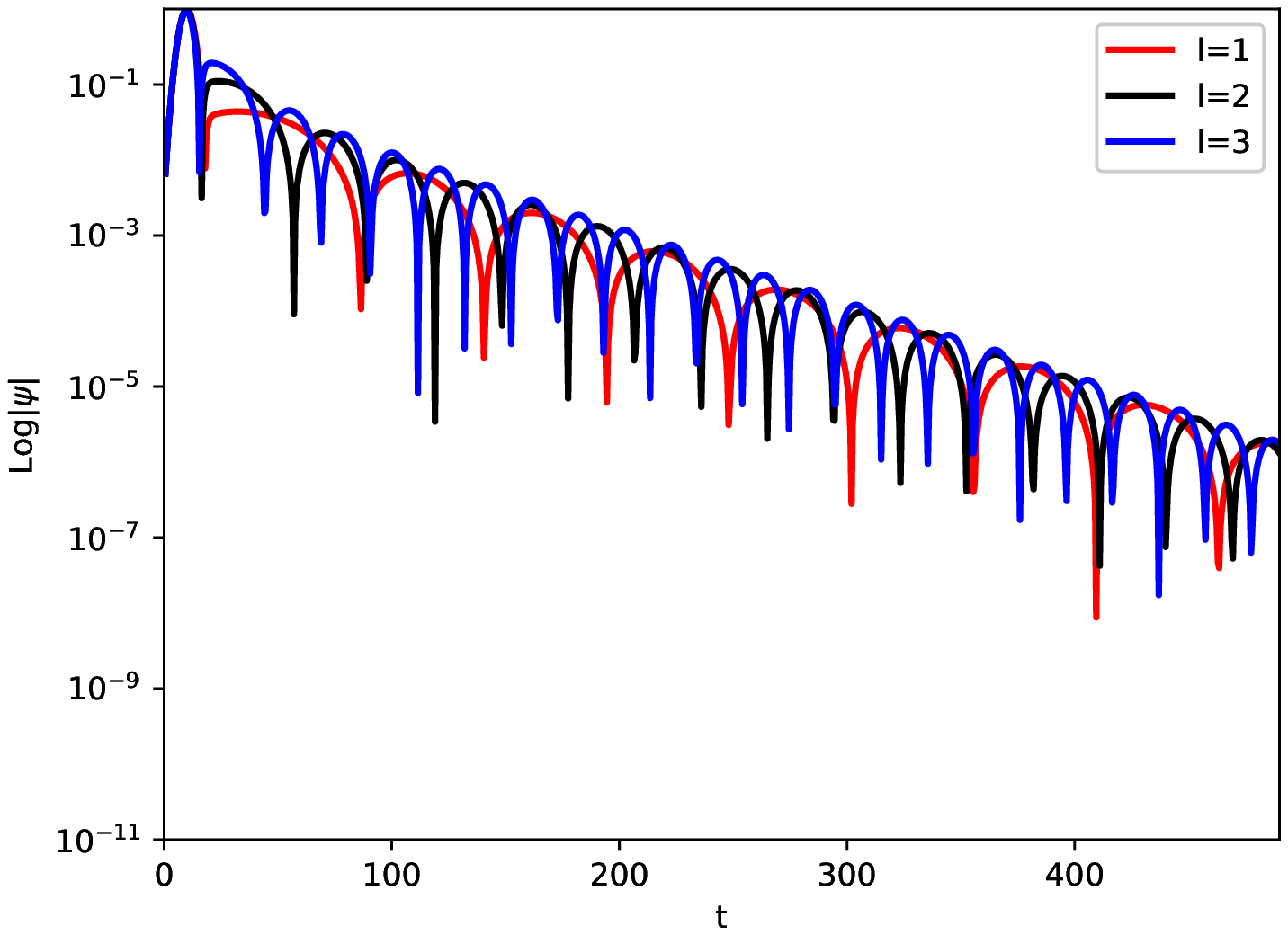}     
} 
\subfigure[]{ 
\label{fig:b}     
\includegraphics[width=0.3\columnwidth]{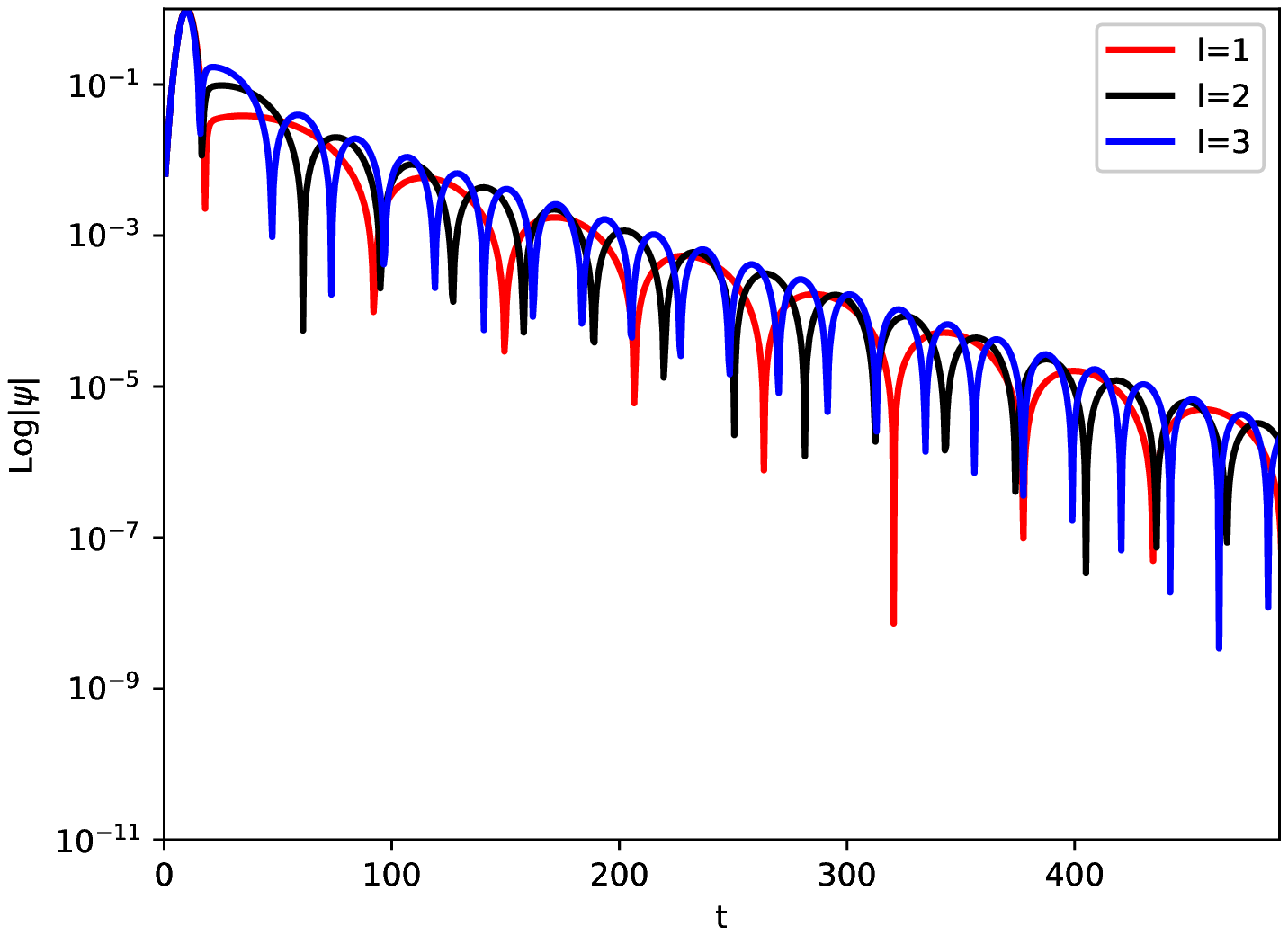}     
}
\subfigure[]{ 
\label{fig:b}     
\includegraphics[width=0.3\columnwidth]{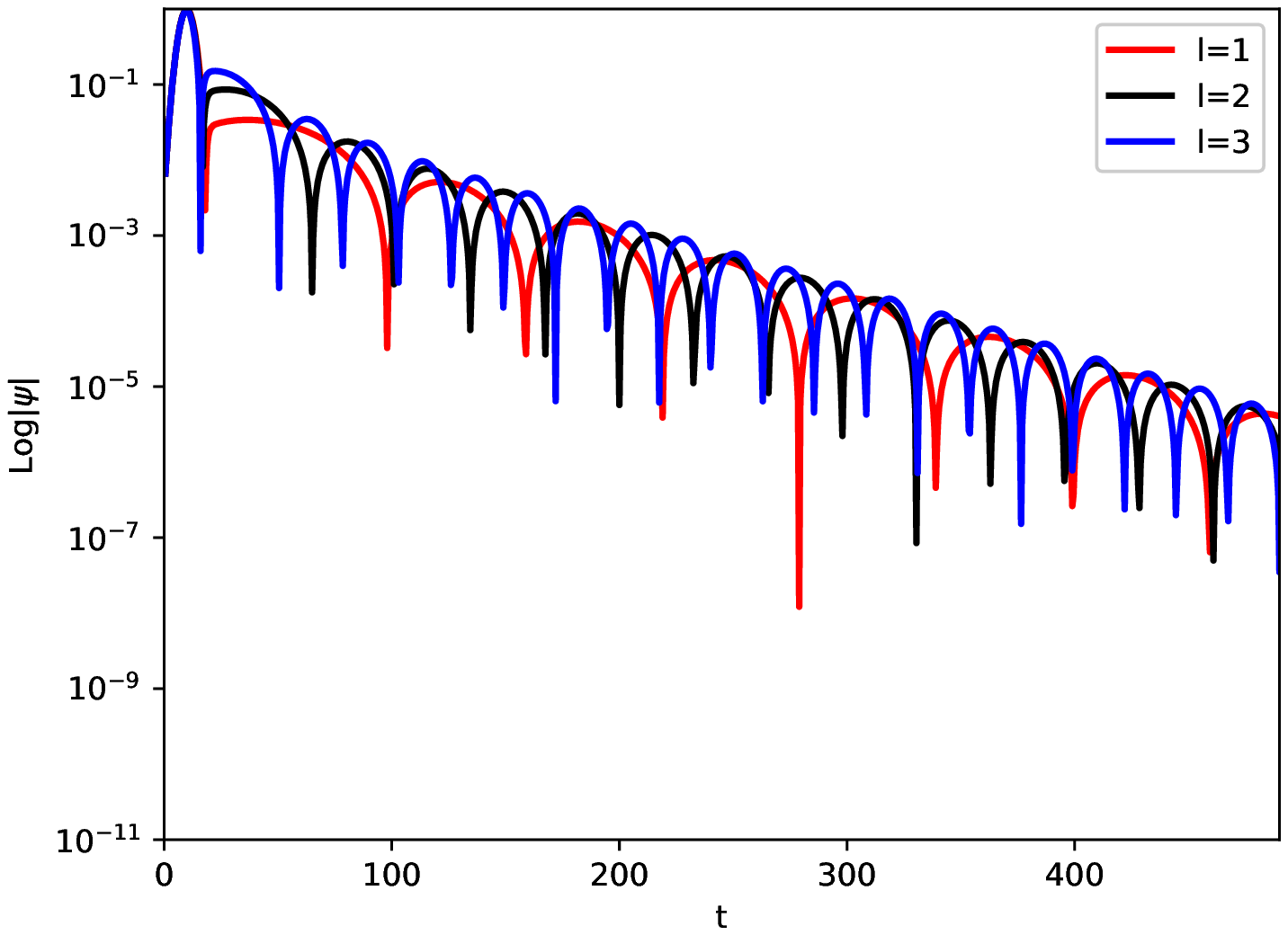}     
}
\caption{The dynamical evolutions of the electromagnetic field with the different $\beta$. (a) $\beta=0.5$ (b) $\beta=1.0$ (c) $\beta=1.5$.}     
\label{fig:8}
\end{figure*}

\begin{figure*}[t!]
\centering
\subfigure[]{ 
\label{fig:b}     
\includegraphics[width=0.4\columnwidth]{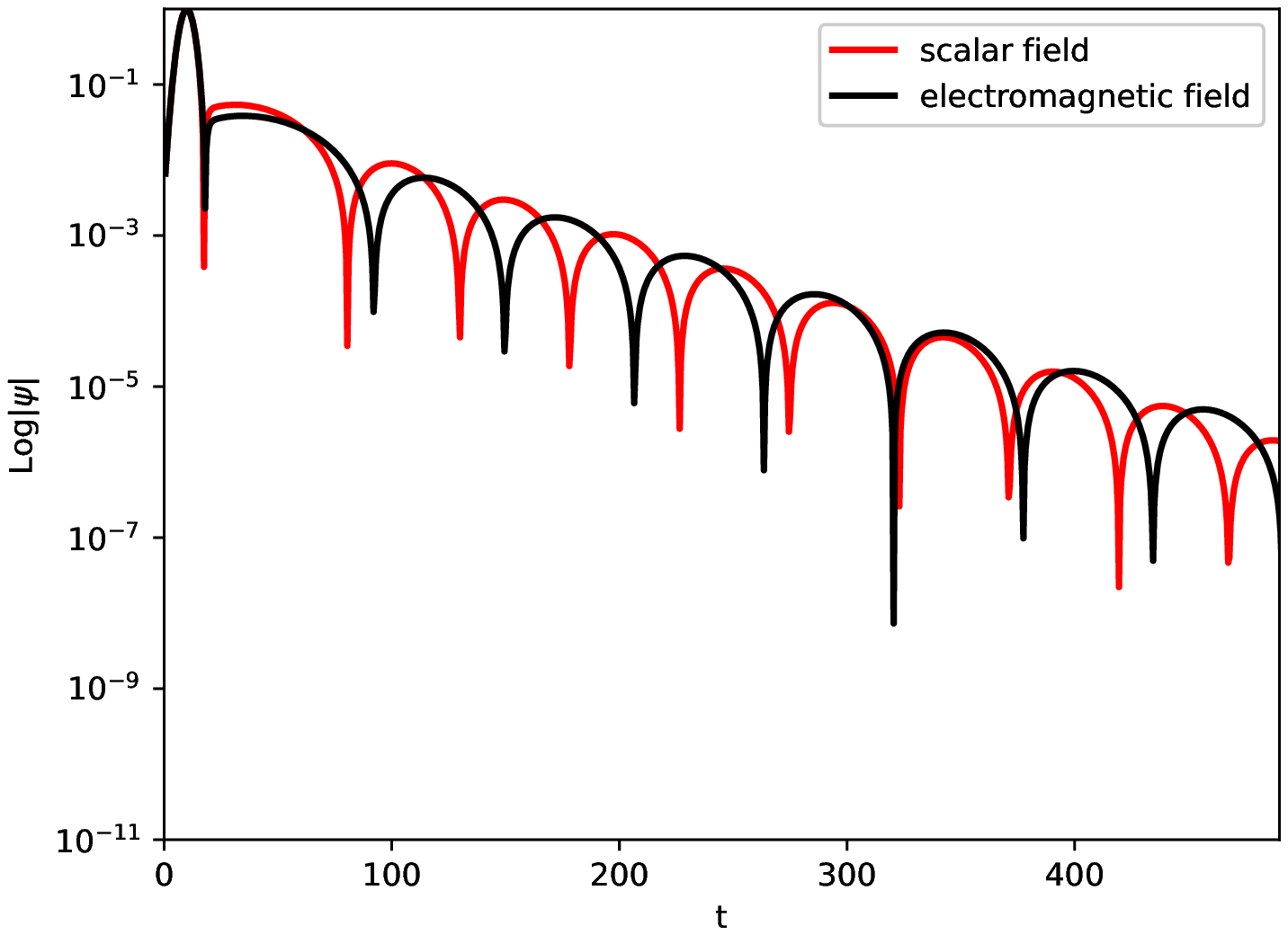}     
} 
\subfigure[]{ 
\label{fig:b}     
\includegraphics[width=0.4\columnwidth]{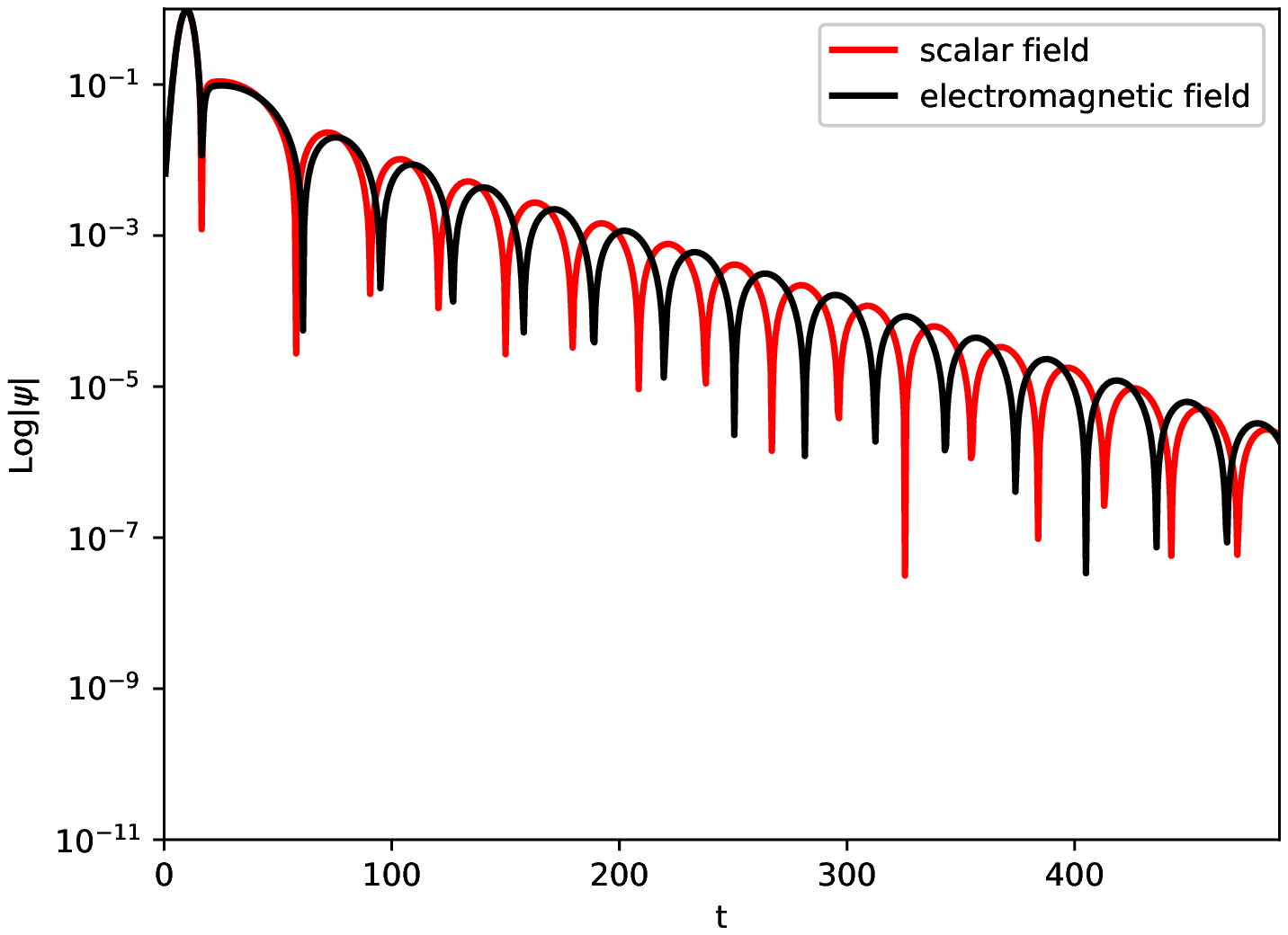}     
}

\caption{The dynamical evolutions of the scalar and electromagnetic field. (a) $\beta=1, l=1$ (b) $\beta=1,l=2$.}     
\label{fig:8.1}
\end{figure*}

\section{Quantum-corrected Schwarzschild black hole model} 
\label{sec:3}
\indent In this section, we discuss the variation law of QNM for scalar and electromagnetic fields. To obtain the law we analyze the GUP corrected Schwarzschild black hole degree gauge, where the solution to Eq.(\ref{equ15}) reflects the evolution process of the QNM. We respectively take $M=1, M_\text{P1}=0.5$, the different parameters of $l$ and $\beta$, and then find the QNM variation law. We know that the variation law of the QNM for the modified Schwarzschild black hole is related to the effective potential. Since we want to quantitatively analyze this qualitative action, we have worked out the QNM frequencies via the sixth-order WKB method in the presence of scalar and electromagnetic fields. According to Tables \ref{tab:1}-\ref{tab:2}, it can be seen that the corrected Schwarzschild black hole returns to steady state after a period of perturbation because the real part of the frequency is positive and the imaginary part is negative. As $l$ is the same, the real imaginary parts of QNMs are decreasing as $\beta$ increases, in other words, the oscillation frequency and damping rate are decreasing, and the more $\beta$ the slower the decay. Our results are opposite to the variation pattern of the real and imaginary parts of the QNM of the second kind aether black hole in \cite{ppp1}, whose real and imaginary parts of the QNM decrease with decreasing parameters $c_{13}$. In the scalar field, when $\beta$ is the same, as $l$ increases, the real part of QNM increases and the imaginary part decreases. With larger $l$, the real part grows more rapidly and the imaginary part decays more slowly. In the electromagnetic field, the real and imaginary parts of the QNM increase as $l$ increases. In a physical sense, the introduction of the parameter $\beta$ can transform the singularity construction of the black hole metric since there are singularities in the solution of the black hole. This new GUP of correction items may contribute to the discretization in time and space. 

\indent As can be seen from Figs.\ref{fig:5}-\ref{fig:6}, for $l$ to be a constant value, the period of the QNM perturbation time is the shortest when $\beta = 0.5$, followed by $\beta = 1.0$ and the longest period of the QNM perturbation when $\beta = 1.5$. In Figs.\ref{fig:7}, it can be seen that when $\beta$ is a fixed value, the period of the perturbation time of QNM is the least when $l$ is much larger, and the magnitude of its period time is $l=0>l=1>l=2$. Our results are in agreement with the conclusion of the time-domain diagram of the QNM for black holes in bumblebee gravity \cite{ppp2} under the scalar field that the parameter $l$ affects the oscillation of the QNM, and the larger $l$ is, the easier the QNM is to detect. Similarly, in Figs.\ref{fig:8}, the magnitude of the cycle time of the perturbation time of QNM is $l=1>l=2>l=3$.  Figs.\ref{fig:8.1} show that if the parameters $l$ and $\beta$ are fixed, the QNM of the scalar field is greater than the effective potential of the electromagnetic field. The average value of the ringing of the QNM in the scalar and electromagnetic fields is about $10^{-1}$ and $10^{-8}$. We can observe that the perturbation process in Fig.\ref{fig:b1} is clearly different from Fig.\ref{fig:b2} and Fig.\ref{fig:b3}, due to the fact that the parameter $l$ affects the frequency of the QNM.

\section{Conclusions}\label{sec:4}
\indent In the present work, we study the QNM of the Schwarzschild black hole corrected by the GUP. We analyze the equations in scalar and electromagnetic fields and obtain the effective potential which corresponds to them. The calculation is performed by the sixth-order WKB method, which gives the frequency of the QNM. We mainly obtained the following results: 

(1)The effective potential of the GUP quantum-corrected Schwarzschild black hole decrease with the increase of $\beta$, and increase with the increase of $l$.  When $r_*$ approaches infinity, the effective potential tends to $0$ and the black hole will be unaffected by the change of parameters and finally comes back to the equilibrium state.

(2)The QNM ringing of the GUP quantum-corrected Schwarzschild black hole appears after the initial pulse. When $l$ is fixed, the QNM oscillation becomes weaker as $\beta$ increases, the correction term of GUP makes the black hole metric gauge not singular; when $\beta$ is fixed, the QNM oscillation becomes stronger as $l$ increases. It can be concluded that the trend of QNM is related to the effective potential.

(3)The QNM is a special characteristic of the solution of the black hole perturbation equation, which occupies the main time in the black hole perturbation process. The parameter $l$ influences the perturbation time of QNM, among the most obvious in Fig.\ref{fig:b1}, which affects the spacetime with black hole background. In other words, it affects the intrinsic frequency as well as the damping rate of the oscillation and is not affected by the initial oscillation.

(4)In Figs.\ref{fig:5}-\ref{fig:6}, when $l$ is a constant value, the period of the time of QNM perturbation is $\beta=1.5>\beta=1.0>\beta=0.5$, then $\beta=0.5$ is easier to detect. In Fig.\ref{fig:7}, it is known that whenever $\beta$ is a constant value, the duration of the perturbation time is $l=0>l=1>l=2$, then $l=2$ is more easily to be detected. In Fig.\ref{fig:8}, it is well recognized that as long as $\beta$ is a constant, the length of the perturbation time is $l=1>l=2>l=3$, then $l=3$ is easy to be found.

(5)The average value of the ringing of QNM in scalar and electromagnetic fields is about in the range of $10^{-1}$ and $10^{-8}$, which is in agreement with our results in Tables \ref{tab:1}-\ref{tab:2} .

(6)The QNM ringing of the scalar is larger than that of the  electromagneticfield field, which means that the radiation excited by the perturbation in the scalar field may be larger than that excited by the electromagneticfield field.

\indent In the present work, we have taken into account the scalar and electromagnetic field perturbations, and it is known that the QNM signals of both perturbations have very close physical effects in the vicinity of the black hole horizon. It is of significance that the gravitational field perturbation is also very interesting, and according to the study \cite{40,41,42,43}, the QNM signal can also be produced under gravitational perturbation. More gravitational radiation is produced under the gravitational field than under the outfield perturbation. Similarly we think it is also an interesting topic to consider quantum corrections below Schwarzschild black holes under gravitational fields.


\end{CJK*}  
\end{document}